\documentclass[12pt,preprint]{aastex}
\def\lsim{\raise0.3ex\hbox{$<$}\kern-0.75em{\lower0.65ex\hbox{$\sim$}}}
\def\gsim{\raise0.3ex\hbox{$>$}\kern-0.75em{\lower0.65ex\hbox{$\sim$}}}

\begin{document}

\title{Protostellar Turbulence Driven by Collimated Outflows} 
 
\author{Fumitaka Nakamura\altaffilmark{1,2} and Zhi-Yun Li\altaffilmark{3}} 

\altaffiltext{1}{Faculty of Education and Human Sciences, Niigata University,
8050 Ikarashi-2, Niigata 950-2181, Japan; fnakamur@ed.niigata-u.ac.jp}
\altaffiltext{2}{Visiting Astronomer, Division of Theoretical
Astrophysics,
National Astronomical Observatory, Mitaka, Tokyo 181-8588, Japan}
\altaffiltext{3}{Department of Astronomy, University of Virginia,
P. O. Box 400325, Charlottesville, VA 22904; zl4h@virginia.edu}

\begin{abstract}

The majority of stars are thought to form in clusters. Cluster formation 
in dense clumps of molecular clouds is strongly influenced, perhaps 
controlled, by supersonic turbulence. We have previously shown that 
the turbulence in regions of active cluster formation is quickly 
transformed by the forming stars through protostellar outflows, and 
that the outflow-driven protostellar turbulence is the environment 
in which most of the cluster members form. Here, we take initial steps 
in quantifying the global properties of the protostellar turbulence 
through 3D MHD simulations. We find that collimated outflows are 
more efficient in driving turbulence than spherical outflows that 
carry the same amounts of momentum. This 
is because collimated outflows can propagate farther away from 
their sources, effectively increasing the turbulence driving 
length; turbulence driven on a larger scale is known to decay 
more slowly. Gravity plays an important role in shaping the 
turbulence, generating infall motions in the cluster forming
region that more or less balance the outward motions driven by 
outflows. The resulting 
quasi-equilibrium state is maintained through a slow rate of star 
formation, with a fraction of the total mass converted into 
stars per free fall time as low as a few percent. 
Magnetic fields are dynamically important even in magnetically
supercritical clumps, provided that their initial strengths are 
not far below the critical value for static cloud support. 
They contain an energy comparable to the turbulent energy of 
the bulk cloud material, and can significantly reduce the rate of 
star formation. We find that the mass 
weighted probability distribution function (PDF) of the volume 
density of the protostellar turbulence is often, although not 
always, approximately lognormal. The PDFs of the column density  
deviate more strongly from lognormal distributions. There is 
a prominent break in the velocity power spectrum of the 
protostellar turbulence, which may provide a way to distinguish 
it from other types of turbulence. 

\end{abstract}

\keywords{ISM: clouds --- ISM:magnetic fields --- MHD --- stars: formation
--- turbulence}

\section{Introduction}
\label{intro}

We investigate the role of protostellar outflows in cluster formation. The 
formation 
of stellar clusters is important to study for at least two reasons. 
First, the majority of stars are thought to form in clusters (Lada \& 
Lada 2003; Allen et al. 2006). This follows from the observations that 
most stars are born in giant molecular clouds (GMCs) and that in nearby 
GMCs such as the Orion molecular clouds 
where systematic surveys are available the majority of young stellar 
objects are found in clusters. Just as importantly, it is in clusters 
that most, if not all, massive stars are produced. Understanding 
cluster formation is likely a prerequisite to understanding massive 
star formation. 

Outflows play an important role in star formation in general and
cluster formation in particular. Individually, they may drive 
the accretion in the inner disk (K\"onigl \& Pudritz 2000), brake 
the stellar rotation (Shu et al. 2000), and perhaps help defining the 
most fundamental quantity of a star---the mass (Nakano et al. 1995; 
Matzner \& McKee 2000; Shu et al. 2004). Collectively, they 
have the potential to replenish the energy and momentum dissipated 
in a star-forming cloud. This possibility was first examined in 
detail by Norman \& 
Silk (1980). They envisioned the star-forming clouds to be constantly 
stirred up by the winds of optically revealed T Tauri stars. The 
idea was strengthened by the discovery of molecular outflows (McKee
1989), which point to even more powerful outflows from the stellar 
vicinity during the embedded, {\it protostellar} phase of star 
formation (Lada 1985; Bontemps et al. 1996). Shu et 
al. (1999) estimated the momentum output from protostellar outflows 
based on the Galactic star formation rate, and concluded that it is 
sufficient to sustain in all molecular gas a level of turbulence at 
$\sim 1 - 2$ km/s, not far from the line widths actually observed 
in typical GMCs. If the majority of stars are formed in localized 
parsec-scale dense clumps that occupy a small fraction of the GMC 
volume (Lada et al. 1991), their ability to influence the dynamics 
of the bulk of the GMC material will probably be reduced; other 
means of turbulence maintenance may be needed in regions of 
relatively little star formation, as concluded by Walawender 
et al. (2005) in the case of the Perseus molecular cloud. The 
concentration of star formation should, however, make the 
outflows more important in the spatially limited, but arguably 
the most interesting, regions of a GMC---the regions of cluster 
formation, where the majority of stars are thought to form. 

We have made a start in simulating cluster formation in magnetized
parsec-scale dense clumps including outflows (Li \& Nakamura 2006). 
We find that, in agreement with previous work (Mac Low et al. 1998; 
Stone et al. 1998; Padoan \& Nordlund 1999), the initial  
turbulence that the clump inherits from its 
formation process decays away quickly. It is replaced by the motions 
driven by the protostellar outflows associated with star formation. 
It is in this protostellar outflow-driven turbulence (``protostellar 
turbulence'' for short hereafter) that the majority of the cluster 
members are produced. In this picture, quantifying the protostellar 
turbulence becomes a pressing issue. The current investigation is 
a step in this direction. 

An important issue that we seek to address is how fast stars 
form in a cluster. The star formation rate (SFR) can be 
constrained by observations. 
For example, Lada et al. (1996) derived an average SFR of 
${\dot M}_*\approx 4.5\times 10^{-5}$~M$_\odot$~yr$^{-1}$ for the 
nearby embedded cluster associated with the reflection nebula 
NGC 1333 in the Perseus molecular cloud, based 
on an estimate of the total stellar mass ($\sim 45$ M$_\odot$) and 
the duration of star formation ($\sim 10^6$ years; see also Aspin 
2003). It is to be compared 
with the limiting rate expected in the case of unimpeded star 
formation (where most of the cloud mass is converted into stars 
in one free-fall time) 
\begin{equation} 
{\dot M}_{\rm ff}= {M\over {\bar t}_{\rm ff}} = \left({32 G\over 3\pi}
\right)^{1/2} \left({M\over L}\right)^{3/2} = 3.91 \times 10^{-3} 
\left({M\over 10^3 M_\odot}\right)^{3/2} \left({1 {\rm pc} \over 
L}\right)^{3/2} M_\odot {\rm yr}^{-1},
\label{sfr_ff}  
\end{equation} 
where $M$ and $L$ are the mass and size of the region, and ${\bar 
t}_{\rm ff}$ is the free-fall time at the average density ${\bar 
\rho}=M/L^3$. In the case of NGC 1333, Warin et al. (1996) carried
out a detailed mapping of the molecular gas in $^{13}$CO and 
C$^{18}$O, and estimated a mass of 2900~M$_\odot$ from $^{13}$CO
and 950~M$_\odot$ from C$^{18}$O in an area of 650 arcmin$^2$,
corresponding to a size $L=2.60$~pc for an adopted distance of
350~pc. The two masses yield ${\dot M}_{\rm ff}=4.62\times 10^{-3}$ 
and $8.66\times 10^{-4}$~M$_\odot$~yr$^{-1}$, respectively. For
a smaller core region of 140 arcmin$^2$ in area (or $L=1.20$~pc),
a mass of 450~M$_\odot$ is estimated from C$^{18}$O, yielding    
${\dot M}_{\rm ff}=8.93\times 10^{-4}$ M$_\odot$ yr$^{-1}$. Note 
that the values of SFR from C$^{18}$O are comparable on the two 
scales; the decrease in mass for the core is nearly offset by a 
decrease in size. These three values of ${\dot M}_{\rm ff}$ are 
some 20-100 times larger than the actual SFR inferred by Lada et 
al. (1996). Unless most stars in the NGC 1333 cluster form 
within a period of time much shorter than $10^6$ years (which is 
not supported by the presence of relatively evolved Class II YSOs  
and the detailed analyses of Lada et al. [1996] and Aspin [2003], 
although a mini-burst of star formation involving a fraction 
of the stars, particularly Class 0 objects, appears to be ongoing 
in this deeply embedded cluster [Bally et al. 1996]), the SFR must 
be reduced by at least an order of magnitude below the free-fall 
value in order to match observations. This conclusion is 
strengthened by the recent HCO$^+$ and N$_2$H$^+$ observations 
of Walsh et al. (2006), which yield a cluster-wide mass accretion 
rate of order $10^{-4}$~M$_\odot$~yr$^{-1}$. The rate is 
comparable to the star formation rate ${\dot M}_*$ and much 
lower than the free-fall rate ${\dot M}_{\rm ff}$. 

The ratio of the actual to limiting star formation rates, ${\dot 
M}_*/{\dot M}_{\rm ff}$, has a simple physical meaning: the 
fraction of the cloud mass turned into stars in one free-fall 
time. It is the dimensionless star formation rate 
${\rm SFR}_{\rm ff}$ defined in Krumholz \& McKee 
(2005) and Krumholz \& Tan (2006). The latter authors estimated 
values of a few percent for ${\rm SFR}_{\rm ff}={\dot M}_*/{\dot 
M}_{\rm ff}$ in three types of objects of different characteristic 
densities, including giant molecular clouds, infrared dark clouds, 
and Galactic and extragalactic dense gas traced by HCN. The 
values, while uncertain, are compatible with the range that we 
estimated above ($\sim 1\%-5\%$) for NGC 1333. The small values 
for SFR$_{\rm ff}$ imply that star formation does not proceed 
at the limiting free-fall rate in molecular clouds in general 
(Zuckerman \& Evans 1974) and localized cluster-forming regions 
in particular. It must be retarded somehow. 

Cluster formation may be retarded by magnetic fields. The magnetic 
field is particularly well observed in OMC1---the 
nearest region of ongoing massive star formation. It has a 
hourglass shape (produced perhaps by gravitational contraction; 
Schleuning 1998; Houde et al. 2004), a line-of-sight field 
strength of $360\pm 80\ \mu$G from CN Zeeman measurements
(Crutcher 1999). If we adopt an inclination angle of the field 
line of $\vartheta=65^\circ$ to the line of sight determined by 
Houde et al. (2004), the total field strength would become 
$\sim 850 \mu$G, which would imply a global magnetic flux-to-mass 
ratio $\sim 2.6$ times below the critical value for cloud 
support by a static magnetic field. The ratio was estimated based 
on the column density measured along the line of sight, which 
should be higher than the column density along the minor axis 
of the (elongated) clump. Correcting for this projection 
effect would bring the intrinsic flux-to-mass ratio closer to 
the critical value. Crutcher (2005) reviewed the available 
observational data on other regions of cluster/massive star 
formation and concluded that their magnetic fields, when detected, 
appear to have strengths not far from the critical values as well. 
We are thus motivated to include in our simulations moderately 
strong magnetic fields, with strengths up to half of the critical 
value.  We do not consider magnetically subcritical clouds, where 
star formation is enabled by ambipolar diffusion, a process not 
treated in the present paper.

Supersonic turbulence can also slow down (global) star formation. A 
major concern is that the turbulence tends to decay quickly, even in 
the presence of a strong magnetic field (e.g., Mac Low et al. 1998; 
Stone et al. 1998; Padoan \& Nordlund 1999). Unless the cloud is 
short-lived, the observed supersonic turbulence must be replenished. 
A common practice is to drive the turbulence in Fourier space. The 
extent to which such a treatment is adequate for the turbulence 
observed in molecular clouds remains to be ascertained (e.g., 
Elmegreen \& Scalo 2004). In particular, it is unclear whether 
the treatment is applicable to the turbulence in regions of active 
cluster formation, which we have argued previously has a unique 
origin: it is driven from within by protostellar outflows 
accompanying star formation (Li \& Nakamura 2006; see also Cunningham 
et al. 2006 and Matzner 2007). The goal of the paper is to quantify 
this outflow-driven turbulence and the rate of star formation in 
it through numerical simulation. 

The rest of the paper is organized as follows. In \S~2, we describe 
the formulation of the physical model, including the simulation 
setup and numerical code used. The results of the simulations are 
presented in \S~3 and \S~4. We discuss the results and their 
implications in \S~5, and summarize our main conclusions in \S~6. 

\section{Model Formulation}
\label{setup}

\subsection{Initial and Boundary Conditions}

We consider a self-gravitating cloud in a cubic box of length $L$ 
on each side. Standard periodic conditions are imposed at the 
boundaries. 
As in Li \& Nakamura (2006), we choose an initial density 
distribution that is centrally condensed. The distribution is 
motivated by the fact that embedded clusters, such as NGC 1333 
(Ridge et al. 2003) and the Serpens core (Olmi \& Testi 2002), 
are often surrounded by envelopes with density decreasing away
from the cluster. Compared with the oft-used uniform density 
distribution, it has the advantage of better isolating the central
region of active cluster formation gravitationally from its mirror 
images introduced by the periodic boundary condition. The adopted 
functional form for the density profile is 
\begin{equation}
\rho (r) = \left\{
\begin{array}{ll}
\frac{\displaystyle \rho_0}{\displaystyle  1+(r/r_0)^2}, &  r \le r_e=L/2\\
\frac{\displaystyle  \rho_0}{\displaystyle 1+(r_e/r_0)^2}, & r > r_e=L/2 
\end{array} \right. 
\label{density}
\end{equation}
where $\rho_0$ is the peak density at the center and $r$ the spherical 
radius. The two characteristic radii, $r_0$ and $r_e$, measure 
the sizes of the central plateau region and the cloud 
as a whole. We adopt $r_0=r_e/3$, corresponding to a central-to-edge 
density contrast of 10. The density in the region between the cubic 
simulation box 
and the sphere of radius $r_e$ is held constant. 

To ensure that the cloud contains many Jeans masses for fragmentation 
into a cluster, we choose a size for the simulation box that is 9 
times the Jeans length at the cloud center
\begin{equation} 
L_J= \left({\pi c_s^2/G \rho_0}\right)^{1/2}. 
\end{equation}
The isothermal sound speed $c_s=2.66\times 10^4 (T/20 K)^{1/2}$~cm/s, 
where $T$ is the cloud temperature, and the central density $\rho_0 = 
4.68\times 10^{-24} n_{H_2,0}$~g~cm$^{-3}$, with $n_{H_2,0}$ being 
the central number density of molecular hydrogen, assuming 1 He for 
every 10 H atoms. We adopt a fiducial value $L=1.5$~pc, typical 
of the dimensions of cluster-forming clumps. It corresponds to a 
Jeans length $L_J=L/9=0.17$~pc. Scaling the box size $L$ by 1.5~pc, 
we obtain a central number density, 
\begin{equation} 
n_{H_2,0}=2.69 \times 10^4 \left(\frac{T}{20 K}\right) 
\left(\frac{1.5{\rm pc}}{L}\right)^2 {\rm cm}^{-3},
\end{equation}
and a total cloud mass,
\begin{equation} 
M_{\rm tot}=9.39\times 10^2 M_\odot \left(\frac{T}{20 K}\right) 
\left(\frac{L}{1.5{\rm pc}}\right)
\end{equation} 
inside the computation domain. The mass is also 
typical of the nearby cluster-forming clumps for the fiducial 
parameters (e.g., Ridge et al. 2003). 

From the Jeans length and sound speed, we define a gravitational 
collapse time 
\begin{equation}
t_g= {L_J \over c_s} = 6.12 \times 10^5 \left({L\over 1.5{\rm pc}}
\right) \left({ 20 K\over T}\right)^{1/2} ({\rm years}). 
\label{gravtime}
\end{equation}
It is longer than the free fall time at the cloud center
\begin{equation}
t_{\rm ff}=\left({3\pi \over 32 G \rho_0}\right)^{1/2} = 1.88 \times 
10^5 \left({L\over 1.5{\rm pc}} \right) \left({ 20 
K\over T}\right)^{1/2} ({\rm years}), 
\label{freefalltime}
\end{equation}
by a factor of 3.27. The time $t_g$ is $26\%$ longer than the free fall 
time ${\bar t}_{\rm ff}$ at the average cloud density ${\bar \rho}=0.15 
\rho_0$. 

We impose a uniform magnetic field along the $x$ axis at the beginning 
of simulation. The field strength is specified by the parameter 
$\alpha$, the ratio of magnetic to thermal pressure at the cloud 
center, through 
\begin{equation}
B_0=4.73 \times 10^{-5} \alpha^{1/2} \left({T\over 20 K}\right)
\left({1.5{\rm pc}\over L}\right) ({\rm Gauss}). 
\label{fieldstrength}
\end{equation}
In units of the critical value $2\pi G^{1/2}$ (Nakano \& 
Nakamura 1978), the flux-to-mass ratio in the central flux tube is 
\begin{equation}
\Gamma_0 = { 2^{1/2} \alpha^{1/2} \over 3\pi\; \tan^{-1}(r_e/r_0)} 
=0.12 \alpha^{1/2}.
\label{centralgamma}
\end{equation}
The dimensionless flux-to-mass ratio for the cloud as a whole is 
${\bar \Gamma}=0.33 \alpha^{1/2}$, 
which is nearly three times larger than the
central value. The envelope is more strongly magnetized than the
central region.

In addition to the magnetic field, we include in our simulation 
a modest level of rotation. Although slow rotation is generally 
difficult to measure in a strongly turbulent cluster-forming 
environment, it has been claimed in some cases, such as the 
Serpens core (Olmi \& Testi 2002). The inferred level of rotation 
is typically not high enough to prevent the collapse of the 
cloud as a whole. It could become more important on smaller 
scales. We assume that the clump rotates slowly around the $z$ 
axis (perpendicular to the initial magnetic field lines), with 
a profile
\begin{equation}
V_{\rm rot} = \left\{
\begin{array}{ll}
\frac{\displaystyle 3 (\varpi/r_0) c_s} {\displaystyle 
1+(\varpi/r_0)^2}, & \ \ \ \varpi < r_e  \\ 0, & \ \ \ \varpi 
\geq r_e \end{array}\right.
\label{rotation}
\end{equation}
where $\varpi$ is the distance from the rotation axis that passes 
through the center of the simulation box. The rotation speed peaks 
at a cylindrical radius $\varpi=r_0=L/6$, with a maximum value 
of $1.5 c_s$. It is set to zero at the simulation boundaries to 
satisfy the periodic condition. This distribution of rotation 
speed is used in all simulations. 

We stir the cloud at the beginning of the simulation with a turbulent 
velocity field of power spectrum $v_k^2 \propto k^{-3}$ and a mass 
weighted rms Mach number ${\cal M}=10$. Following the standard 
practice (e.g., Ostriker et al. 2001), random realization of the 
power spectrum is done in Fourier space. The turbulence is 
allowed to decay freely, except for the feedback from protostellar 
outflows. We have experimented with other choices for the initial
turbulence, including different random realizations of the same 
power spectrum and different power-laws for the spectrum 
(e.g., $v_k^2\propto k^{-1}, k^{-2}$, and $k^{-4}$), 
and 
obtained qualitatively similar results. In particular, the fully
developed protostellar turbulence is insensitive to the variations
in the initial turbulence. 

\subsection{Prescriptions for Stellar Mass and Outflow}
\label{prescription}

How the mass of a star is determined is a long-standing, unresolved 
problem. Low-mass stars are observed 
to form in dense cores of molecular clouds. There is evidence that 
only a fraction of the core material eventually ends up inside stars. 
For example, Onishi et al. (2002) derived an average virial mass of 
about 5$M_\odot$ for the H$^{13}$CO$^+$ cores in the Taurus molecular 
clouds. This mass is an order of magnitude higher than that of a 
typical star formed in the region, which is only $\sim 0.5~M_\odot$ 
(Kenyon \& Hartmann 1995). Dense 
cores in more crowded cluster-forming regions are more difficult 
to resolve observationally. They may convert a higher fraction of 
their mass into stars (Motte et al. 1998), although this is uncertain. 
In regions far from massive stars, the masses of low-mass stars are 
probably determined by the competition between protostellar infall 
and outflow, neither of which is fully understood. 
In our treatment, we will leave the stellar mass as a quantity to 
be prescribed. 

To specify the stellar mass, we adopt the following recipe for mass
extraction. When the density in a cell crosses a threshold $\rho
_{\rm th}= 100 \rho_0$, we define around it a ``supercell'' that 
includes all cells in direct contact with the central cell, either 
through a surface, a line or a point. The supercell is thus a cubic 
region having 3 cells on each side and 27 cells in total. 
From each of the 27 cells, we extract 20\% of the mass and put it in 
a Lagrangian particle located at the supercell center; the particle 
represents a formed ``star.'' The percentage 
of the mass extracted is chosen to yield a stellar mass $M_*\approx 
0.5~M_\odot$, typical of low mass stars. We have examined the physical 
conditions of the gas in a large number of supercells, and found the 
gas to be self-gravitating and well on the way to star formation. 
At formation, we assume that the star moves with the velocity of 
its host cell. Its subsequent evolution through 
the cluster potential, which includes contributions from both gas 
and stars\footnote{In computing the contribution of a formed star 
(or Lagrangian particle) to the gravitational potential, we spread 
its mass evenly in its host cell to avoid singularity.}, is 
followed numerically.

Following Matzner \& McKee (2000), we assume that each star injects
into the ambient medium a momentum that is proportional to the
stellar mass $M_*$. We denote the proportionality constant by 
$P_*$, and normalize it by $100$~km/s. In the Appendix, we find 
a likely range for the scaled outflow parameter $f=P_*/100$~km/s 
between $\sim 0.1$ and $\sim 1$. We will adopt $f=0.5$ as our 
fiducial value, corresponding to a round number $P_*=50$~km/s, 
which is close to the value 40~km/s adopted by Matzner \& McKee 
(2000) in their semi-analytic model of cluster formation. For 
comparison, we will also discuss in some depth a simulation with 
weaker outflows specified by $P_*=25$~km/s.

We assume that the outflow momentum $M_* P_*$ ejected by a star of
mass $M_*$ is shared instantaneously by the material left in the 
supercell immediately after the stellar mass extraction. This 
prescription is similar to the one used by Allen \& Shu (2000) in 
their simulation of outflow-driven motions in critically magnetized, 
sheet-like GMCs. In our initial study (Li \& Nakamura 2006), we 
let the momentum-carrying material in the supercell move radially 
away from the center, with the same speed in all directions. However, 
there is ample observational evidence and strong theoretical arguments 
for the protostellar outflows being collimated in general (Bachiller 
\& Tafalla 1999; Shu et al. 1995). Here, we go beyond the simplest 
prescription, and adopt a two-component structure for the outflow, 
with each component serving a distinct purpose: a collimated, 
``jet'' component that facilitates the energy and momentum 
transport to large distances, surrounded by a spherical component 
that reverses the local infall after a star is formed. As in Allen 
\& Shu (2000), we pick the direction of the magnetic field in the
central cell as the jet axis. Included in the jet component are all 
cells of the supercell whose centers lie within 30$^\circ$ of the 
axis; the remaining cells are assigned to the spherical component. 
The jet carries a fraction $\eta$ of the total outflow momentum 
$M_* P_*$; the remaining fraction is carried by the spherical 
component. The material in each 
component is driven away from the center radially, with a speed 
given by the momentum divided by the mass in that component. 

The above prescriptions for stellar mass and outflow, while idealized,
serve as a reasonable first approximation in our view. Possible future 
refinements are described in \S~\ref{improvement}. 

\subsection{Numerical Code}

The MHD equations that govern the cloud evolution are solved using a 3D 
MHD code based on an upwind TVD scheme. To ensure that the magnetic 
field is divergence free, we replace the magnetic field $\mbox{\boldmath 
$B$}$ by $\mbox{\boldmath $B$}^{\rm new} = \mbox{\boldmath $B$}-\nabla 
\Phi_B$ (where $\nabla^2 \Phi_B = \nabla\cdot\mbox{\boldmath $B$}$) after 
each time step. The code does not contain any artificial viscosity, 
which allows the shocks to be captured sharply. We have tested the 
code against standard shock tube problems and the problem of point 
explosion. The tests show that our code is second-order accurate in 
both space and time, and 
that it suppresses a numerical instability
that appears just behind a shock front for non-TVD code
like the Lax-Wendroff code 
[see e.g., Fig. 7 of Nakamura et al. (1999) and 
Fig. 7 of Nakamura (2000)].
The 3D code is an extension of the well-tested 2D codes that 
we have used in several previous studies (e.g., Nakamura et al. 1999 
and Li \& Nakamura 2002). 

We solve the Poisson equation for gravitational potential using 
fast Fourier transform, taking advantage of the 
periodic boundary conditions imposed. As is the standard practice, 
we set the $k = 0$ component of the gravitational potential to 
zero in Fourier space. This is equivalent to using 
$\rho-\bar{\rho}$ (where $\bar{\rho}$ is the mean density 
of the cloud), rather than $\rho$, as the source term in the 
Poisson equation. Formed stars are treated separately from the
gas, as Lagrangian particles. Their equations of motion are 
solved using a symplectic method.

In the simulations to be presented in this paper, we adopt a 
uniform grid of $128^3$. This resolution is relatively low 
compared 
to some recent simulations of MHD turbulence in molecular clouds 
(e.g., Li et al. 2004). However, our simulations include a new 
ingredient---protostellar outflows---and are followed to a time 
significantly later than the previous simulations. Both the 
outflow and longer evolution 
exacerbate the time step problem associated with large Alfv\'en 
speeds in rarefied regions. Restricting ourselves to a relatively 
modest resolution allows us to carry out an exploration of 
parameters essential to the problem. Furthermore, our grid does 
contain more than 2 million cells, which enable us to resolve 
the global structure of the outflow-driven turbulence reasonably 
well. We are running a couple of higher resolution simulations 
with $256^3$. Preliminary analyses show that our main results 
do not depend sensitively on resolution. 

\section{Protostellar Outflow-driven Turbulence: the Standard Model}
\label{outflowsection}

A number of parameters are needed to fully specify our simulations. 
In this paper, we will concentrate on three that most directly 
affect the properties of the protostellar outflow-driven turbulence. 
These include the parameters $f$ and $\eta$, which characterize the 
strength and collimation of the outflow respectively, and the ratio 
of magnetic-to-thermal energy at the cloud center $\alpha$, which 
specifies the degree of cloud magnetization. Other parameters, such 
as those specifying the initial mass 
distribution and turbulent velocity field, are kept 
fixed to facilitate model comparison. In this section, we focus on 
a ``standard'' model with $\alpha=2.5$, $f=0.5$ and $\eta=0.75$ 
(Model S0 in Table 1). The choice $\alpha=2.5$ yields a dimensionless 
flux-to-mass ratio of 0.19 (in units of the critical value $2\pi 
G^{1/2}$) in the central flux tube and 0.52 for the cluster-forming 
clump as a whole. The clump is therefore magnetically supercritical, 
although only moderately so for the bulk of the cloud material. The 
choice $f=0.5$ corresponds to a relatively strong outflow, and 
$\eta=0.75$ means that the outflow momentum is dominated by the 
relatively narrow ``jet'' component. The combination of model 
parameters is chosen to yield a relatively low rate of star 
formation. This simulation will serve as the standard against 
which other simulations, listed in Table~1 and to be discussed 
in the next section, are compared. 

\subsection{Star Formation Rate}

\begin{figure}
\plotone{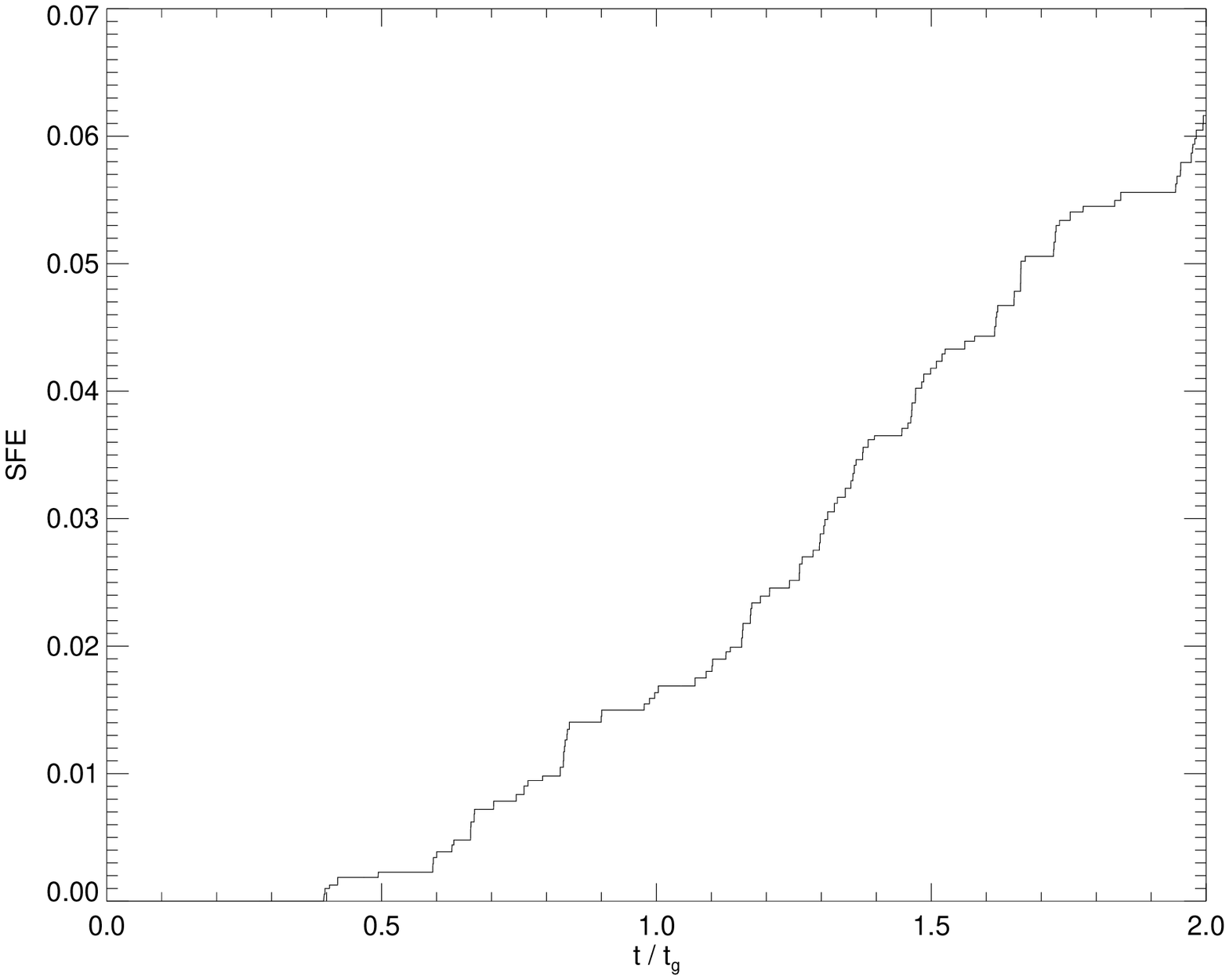}
\caption{Evolution of the star formation efficiency (SFE) in the standard
model (Model S0), showing that only a few percent of the gas is converted 
into stars per free-fall time (${\bar t}_{\rm ff}=0.79 t_g$).}  
\label{SFEstandard}
\end{figure}

One of the most important quantities of cluster formation is the star 
formation efficiency (SFE hereafter), defined as the ratio of the 
mass of all stars to the total mass of stars and gas. It determines, 
among other things, whether a cluster 
would remain gravitationally bound upon gas removal. In 
Fig.~\ref{SFEstandard}, we plot the SFE of the standard model 
as a function of time, in units of the gravitational time $t_g$, 
which is $26\%$ longer than the average free-fall time ${\bar t}
_{\rm ff}$. The first several stars form in quick succession around 
$0.4~t_g$, followed by a period of relative quiescence. Episodes of 
similar mini-bursts of star formation are also evident at later times. 
By the last time shown in Fig.~\ref{SFEstandard} ($t=2~t_g$ or 
$1.22 [L/1.5{\rm pc}][20K/T]^{1/2}$~Myrs), a total of 114 stars 
have formed, with an accumulative SFE of $\sim 6\%$. The average 
mass of a star is $\sim 0.5 \ M_\odot (T/20 K) (L/1.5{\rm pc})$, 
typical of low-mass stars, as mentioned earlier.  
From the fact that within a time interval $\sim 1.6~t_g$ (or $\sim 
2~{\bar t}_{\rm ff}$) since the formation of the first star $\sim 
6\%$ of the gas has 
been converted into stars, we obtain a star formation efficiency 
per free-fall time ${\rm SFR}_{\rm ff}\approx 3\%$. It is in the 
range of star formation rate inferred for the embedded cluster 
NGC 1333 and other types of objects (see discussion in \S~\ref{intro}). 
At this rate, it takes some $33$ free-fall times to completely 
deplete the gas in the cluster forming region through star 
formation\footnote{It does not mean that the clump will live 
for a time as 
long as the depletion time. The clump will likely be dispersed 
either internally or externally long before the conversion of gas 
to stars is completed.}. To understand why the star formation is 
slowed down to such a remarkable extent, we examine in some depth 
the internal structure and dynamics of the cluster-forming clump, 
starting from 
global quantities such as the total energy and momentum. 
The effects of the magnetic field are discussed separately in the 
next section (\S~\ref{magfield}).

\subsection{Evolution of Scalar Momentum and Gravitational Energy}

\begin{figure}
\plotone{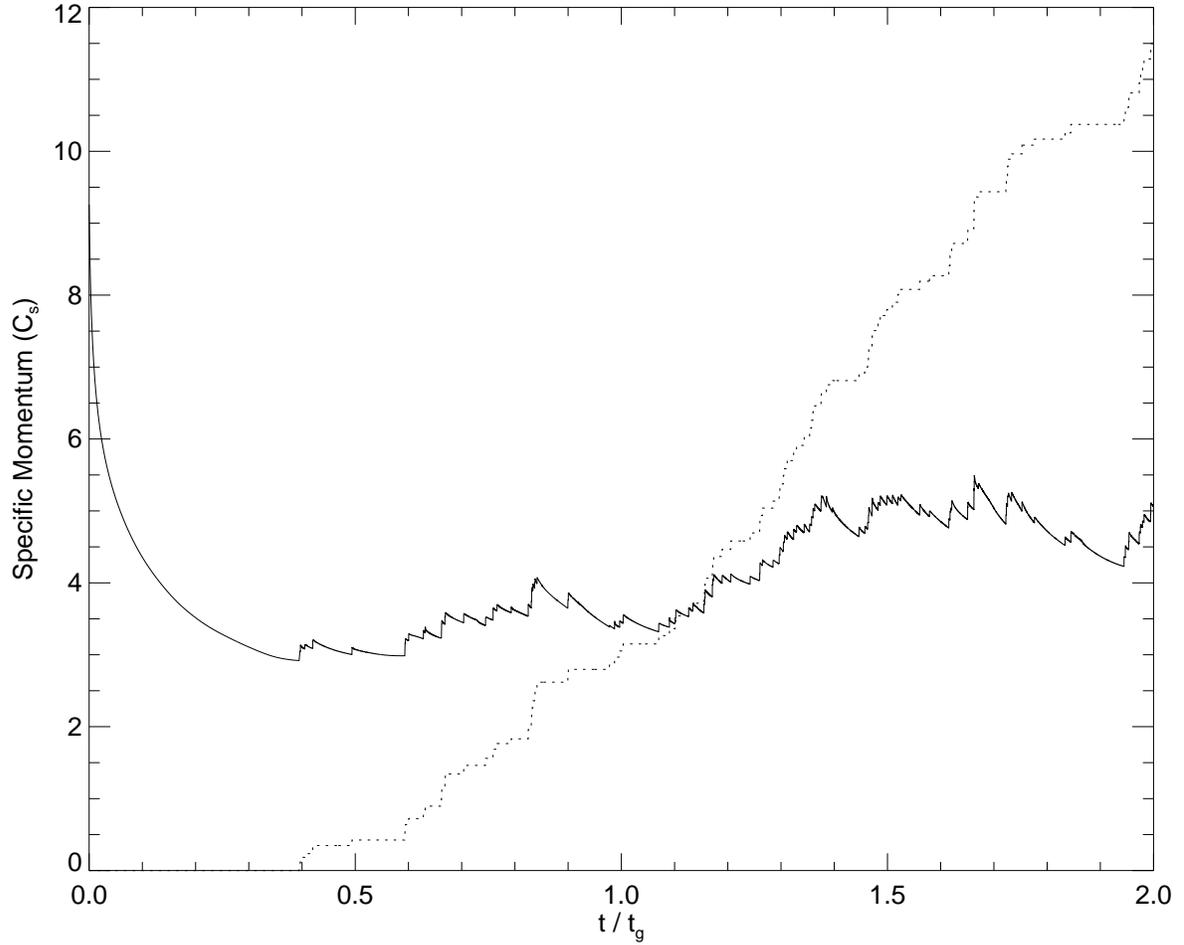}
\caption{Comparison of the specific (scalar) momentum $v_{\rm turb}$ 
(solid line) in the turbulent cluster-forming clump at a given time 
and the amount of specific momentum $v_{\rm inj}$ (dotted) injected 
into the clump by outflows up to that time. }
\label{momentum}  
\end{figure}

The momentum is more useful to keep track of than the kinetic energy 
in our standard simulation. This is because the kinetic energy can 
easily be dominated by the fast moving outflows, especially the jet 
components. Impulsive injection and rapid dissipation of the 
outflow energy generate large variations in the total kinetic energy,
even though the kinetic energy of the bulk cloud material away from
the active outflows remains relatively constant. The total (scalar) 
momentum is not expected to vary as strongly, since the outflows 
interact with the ambient medium in a momentum-conserving fashion. 
In Fig.~\ref{momentum}, we show the evolution of the total (scalar) 
momentum divided by the total mass of the gas. The ratio is the 
specific momentum or simply the mass-weighted turbulent speed, which 
we will denote by $v_{\rm turb}$. It provides a better measure of 
the turbulent speed of the bulk material than the oft-used mass-weighted 
rms velocity; the latter is more dominated by active outflows. As 
expected, the specific momentum decreases rapidly at the beginning 
of the simulation, as a result of the momentum-canceling (head-on) 
collisions that develop in the highly compressive initial turbulent 
velocity field. The decline is stopped by the formation of stars, 
whose outflows appear to have kept the turbulent speed $v_{\rm turb}$ 
close to $\sim5\ c_s$ at late times. The near constancy of $v_{\rm 
turb}$ indicates that the turbulent gas has reached a quasi-equilibrium 
state.

The specific momentum of the turbulent gas is to be compared with the 
specific momentum injected into the gas by outflows. The latter is 
given by $v_{\rm inj}= M_{*,{\rm tot}} P_*/M_{\rm gas}$, where 
$M_{*,{\rm tot}}$ and $M_{\rm gas}$ are the total masses of the 
stars and gas, respectively, and $P_*$ is the outflow momentum 
per unit stellar mass. If only a small fraction of the gas is 
converted into stars, as in the standard simulation, we 
have $v_{\rm inj}\approx {\rm SFE}\times P_*$, where SFE is 
the star formation efficiency. For the adopted $P_*=50$~km/s,
we have $v_{\rm inj} \approx $3~km/s at the end of the simulation 
($2\ t_g$), when ${\rm SFE} \approx 6\%$. It is about twice the 
mass-weighted turbulent speed $v_{\rm turb}\approx 1.5$~km/s 
at the same time. In Fig.~\ref{momentum}, we show the speed 
$v_{\rm inj}$ as a function of time along with the turbulent 
speed $v_{\rm turb}$. From the slope of the $v_{\rm inj}$ curve, 
we estimate the amount of specific momentum injected into the 
cloud by outflows per gravitational time $t_g$ to be $\sim 7\ c_s$. 
At this rate, it takes $\sim0.7\ t_g$ (or $\sim0.9\ {\bar 
t}_{\rm ff}$) to replenish the equilibrium specific momentum 
$v_{\rm turb}\approx 5\ c_s$. In a steady state, this should 
also be the momentum dissipation time. 

\begin{figure}
\plotone{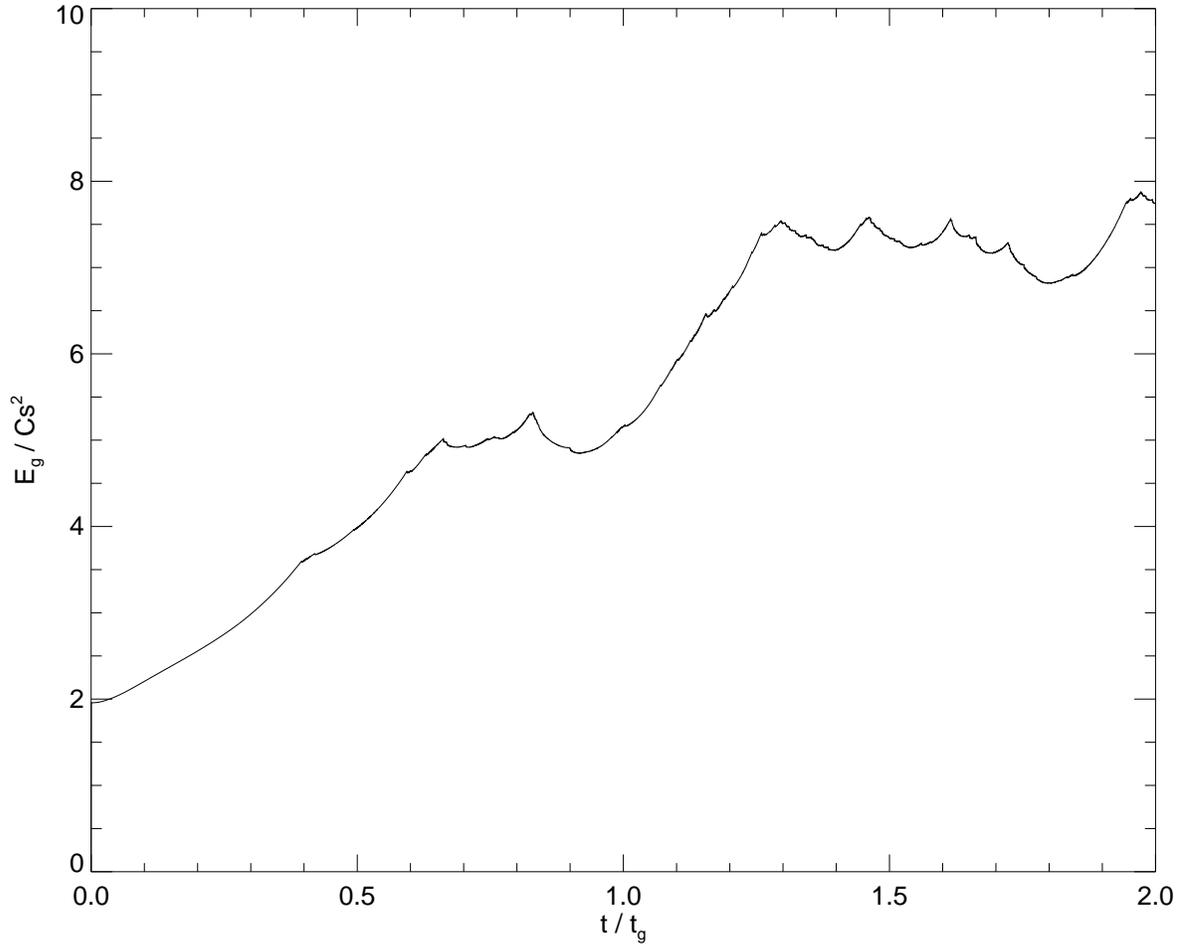}
\caption{Evolution of the absolute value of the gravitational energy per 
unit mass. The energy becomes nearly constant at late times, signaling 
that a quasi-equilibrium state has been reached.}  
\label{gravenergy}
\end{figure}

Another indication that a quasi-equilibrium state is reached at late 
times comes from the evolution of the absolute value of the 
gravitational energy\footnote{We remind the reader that the $k=0$ 
component of the gravitational potential is set to zero in Fourier 
space to accommodate the periodic boundary condition. This treatment 
reduces the absolute value of the gravitational energy in the model 
compared to the case where the cloud exists in isolation.} per unit
mass, $E_g$, shown in Fig.~\ref{gravenergy}. The initial increase 
in $E_g$ is caused by turbulence dissipation, which leads to cloud 
contraction, starting from an initial mass distribution that is 
already somewhat centrally condensed. The contraction is slowed
down, and eventually arrested by the outflows associated with
star formation. The near constancy of the gravitational energy 
at late times implies that the system is neither collapsing 
nor expanding rapidly as a whole. It signals that a quasi-equilibrium
state has been reached. In what follows, we shall examine in 
some detail the structure and dynamics of the 
quasi-equilibrium state, focusing on a representative time 
$1.5~t_g$, when 80 stars have formed and the outflow-driven 
protostellar turbulence appears fully developed.

\subsection{Quasi-Equilibrium State}
\label{decay}

\begin{figure}
\plotone{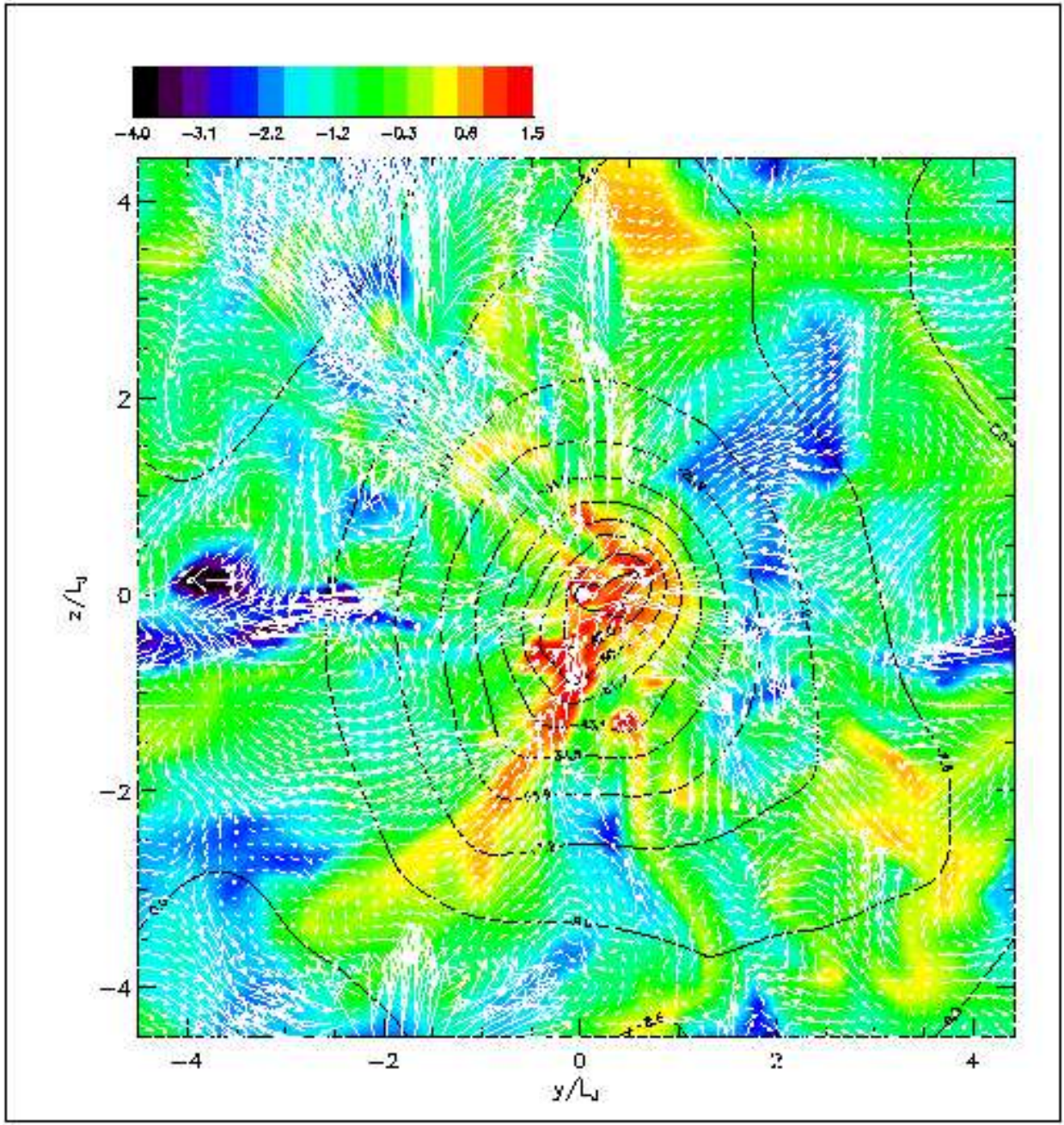}
\caption{A slice through the standard simulation in the $y$-$z$ plane. 
The color map is for the logarithm of the density log$(\rho/\rho_0)$, 
arrows for the velocity vectors and contours for the gravitational 
potential. The contours are plotted at 0.0, 0.1, ..., 0.9 times the 
minimum value of the potential ($-86.2~c_s^2$). For clarity, the 
length of velocity vector is capped at $15\ c_s$ and the location of 
minimum gravitational potential is shifted to the center of the map.}
\label{velvector}
\end{figure}

Global quantities such as the total scalar momentum and gravitational
energy do not capture the full complexity of the gas dynamics. This 
complexity is illustrated in Fig.~\ref{velvector}, where we show a 
color map of the density distribution, with velocity vectors and 
contours of the gravitational potential superposed. Naively, one 
might expect the gas to slide down the gravitational potential well 
towards the bottom more or less freely, producing an 
infall-dominated velocity field that leads to rapid star formation 
at the limiting free-fall rate given by equation~(1). Infall 
motion does appear in some regions, but not in others: many patches 
have more or less coherent motions in directions around or away 
from the bottom of the potential well. The patchy appearance of the
velocity field is a general feature of the protostellar turbulence 
that we also see at other times and in other simulations. The 
patches are often separated by regions of enhanced density, which 
are probably created by converging flows. An important question 
is: are the infall and outflow motions more or less balanced 
globally so as to keep the self-gravitating system close to a 
dynamical equilibrium?

\begin{figure}
\plotone{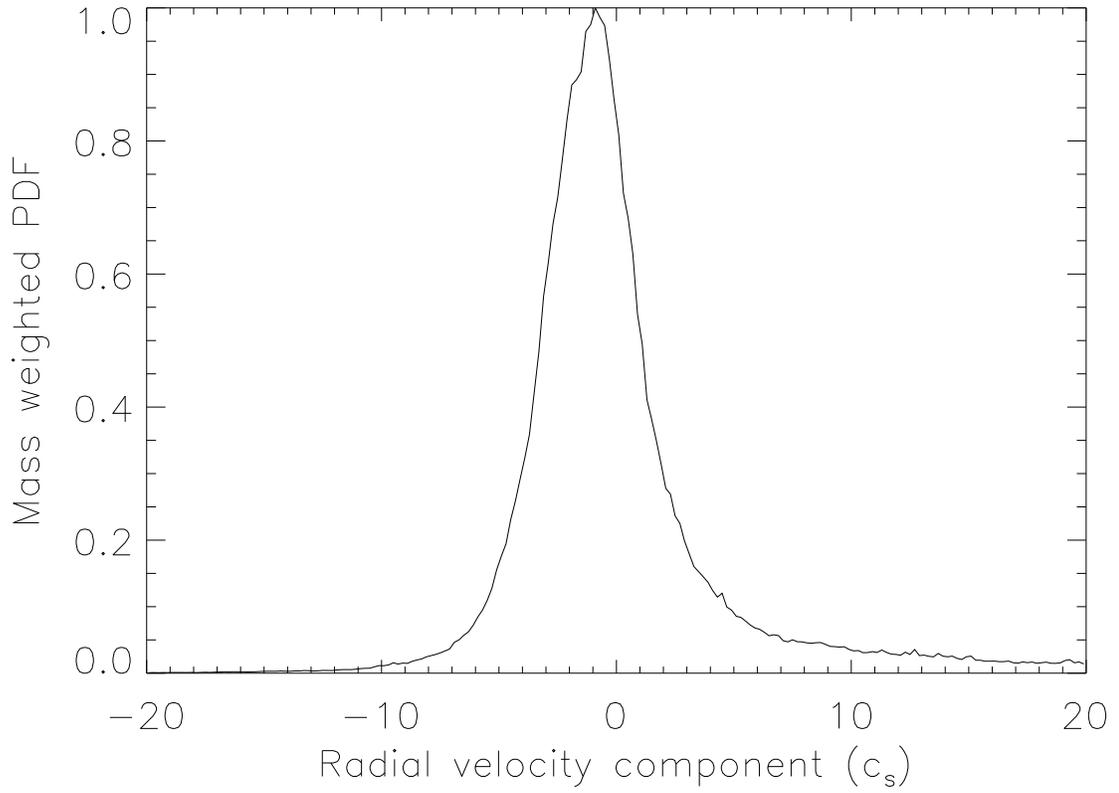}
\caption{Mass weighted PDF for the radial velocity component. The 
maximum of the distribution is normalized to unity.} 
\label{velPDF}
\end{figure}

To address the above question quantitatively, we show in Fig.~\ref{velPDF}
the mass weighted probability distribution function (PDF) for the radial 
component of the velocity, $v_r$, towards or away from the location 
of minimum gravitational potential. The PDF peaks around 
$v_r\approx -1~c_s$, and a simple integration yields that $61\%$ of the 
total mass has a negative radial velocity and is thus infalling. The 
outflowing gas moves somewhat faster on average, however. If we define 
an average infall (outflow) speed from the total inward (outward) momentum 
divided by the total infalling (outflowing) mass, we obtain a value 
of $3.99\ c_s$ for outflow and $-2.40\ c_s$ for infall. The faster
outflow speed is consistent with the fact that the PDF shows a stronger 
wing towards positive $v_r$. The {\it net} radial velocity, defined as 
the {\it net} total radial momentum divided by the total mass, is only 
$0.08\ c_s$, much smaller than either the average infall or outflow 
speed. It is only a small 
fraction of the sound speed, indicating the outflow and infall 
motions are nearly balanced globally, leaving the system in a 
rough, dynamical equilibrium in the radial direction. The balance
is generally true at other times as well, although the net radial
speed can go up to several tenths of the sound speed.

\subsection{Mass Distribution}

\begin{figure}
\epsscale{1.75}
\plottwo{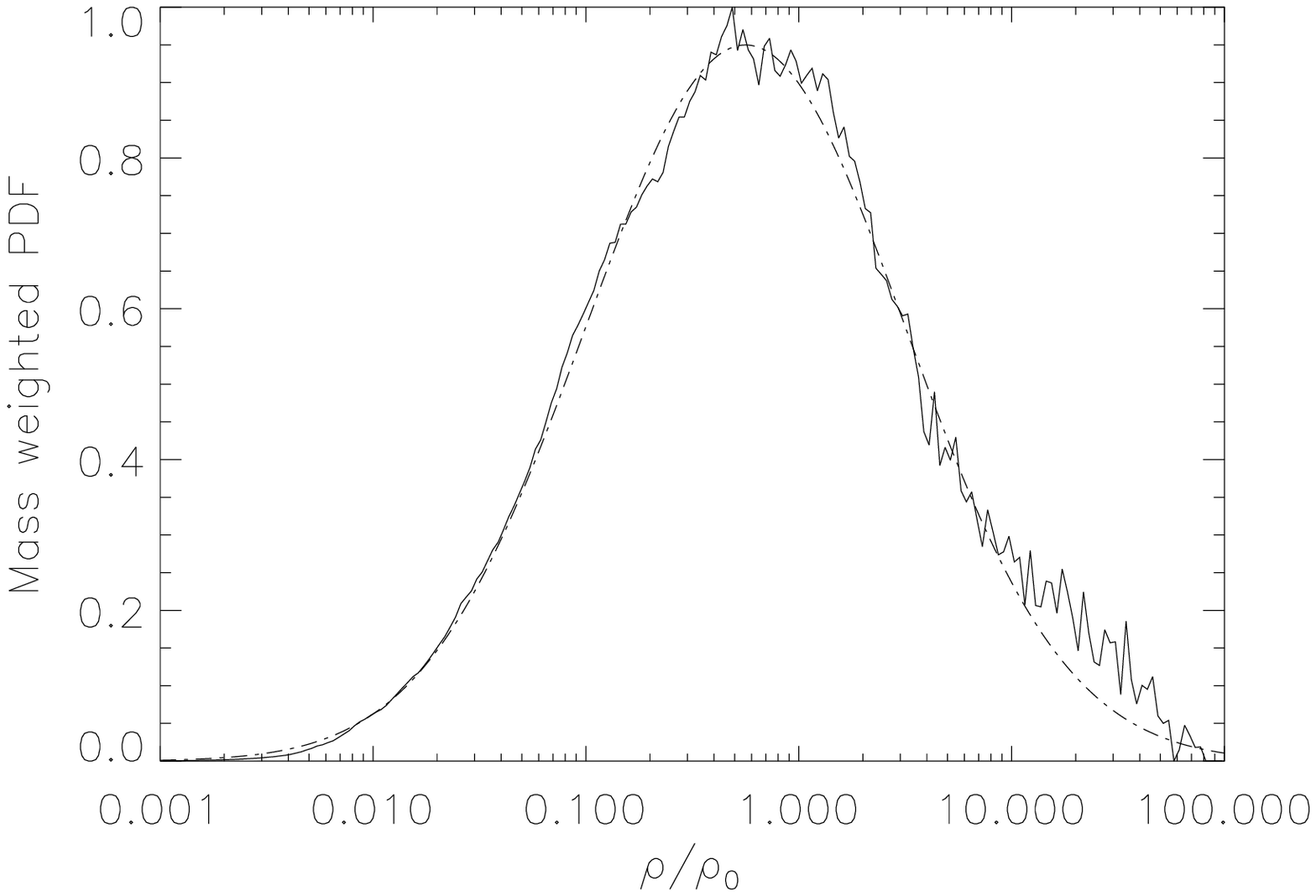}{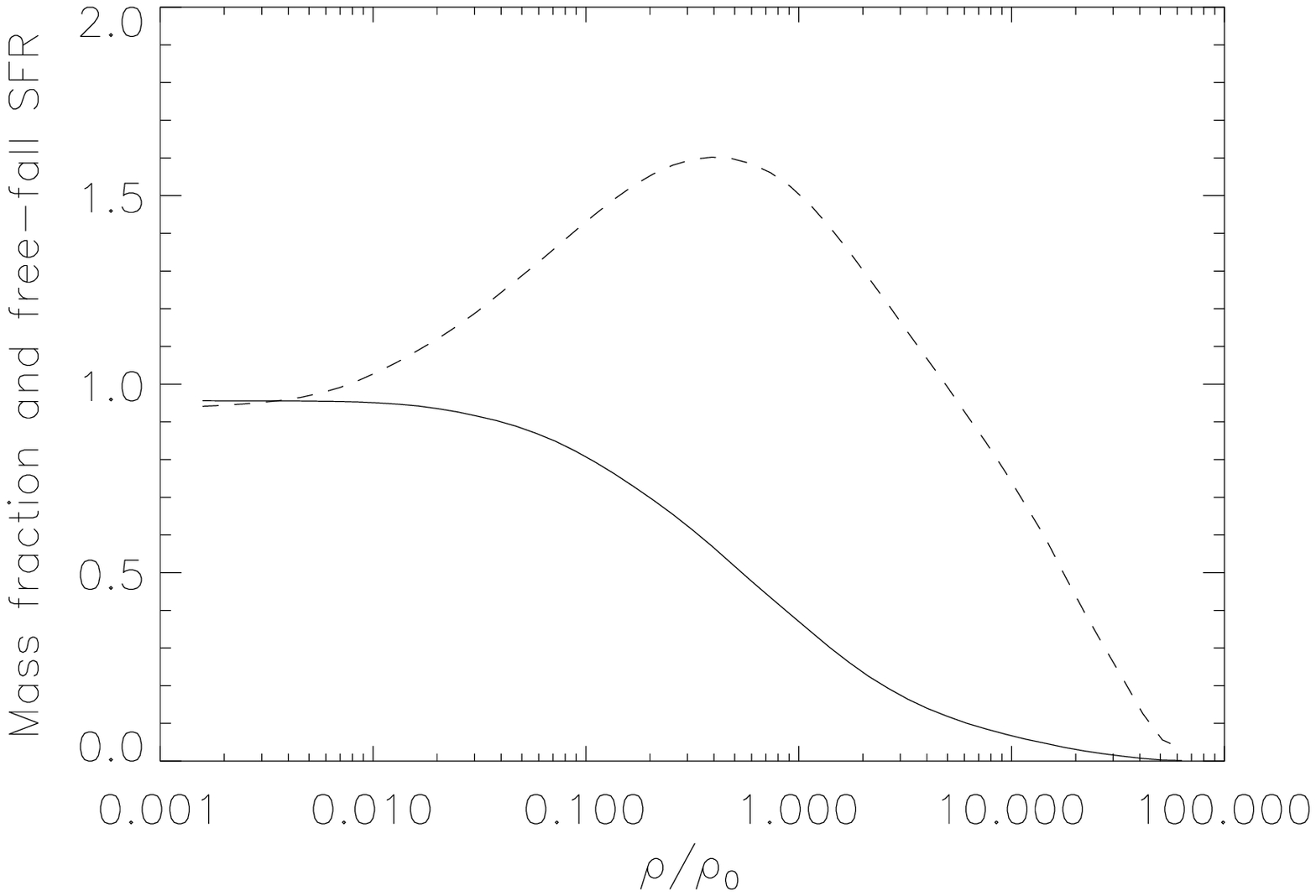}
\caption{(a) Mass weighted probability distribution function (PDF) for the 
volume density (solid line). Also plotted is a fitted lognormal distribution
for comparison. (b) The accumulative mass fraction (solid) and the limiting 
(free-fall) rate of star formation (dashed) for the gas above a given 
density. 
}
\label{SFRff}  
\epsscale{1.0}
\end{figure}

It is well known that supersonically turbulent media are clumpy, 
with a wide range of densities. In Fig.~\ref{SFRff}(a), we plot 
the mass weighted PDF for the volume density at the representative 
time. The distribution appears roughly lognormal, peaking 
around $\sim 0.6~\rho_0$ (or $\sim 4$ times the average density), 
broadly consistent with previous simulations (e.g., Ostriker et 
al. 2001), despite the fact that in our simulation the supersonic 
turbulence is created and maintained in a potential well of  
considerable depth and gravity plays a more dominant role. There 
is some deviation at the high density end, where the PDF is 
dominated by dense cores and filaments, some of which are 
self-gravitating and are on the verge of collapse to form a 
new generation of stars. In Fig.~\ref{SFRff}(b), we plot 
the fraction of the total mass that resides in regions above 
a given density. At low densities, the mass fraction asymptotes 
to a value slightly below unity, because 
a few percent of the mass has already been converted into stars 
at the time under consideration. It decreases quickly towards 
the high density end. Only $12\%$ of the mass resides in regions 
denser than $5~\rho_0$, and this fraction drops to $5.8\%$ 
above $10~\rho_0$. The rapid decrease in the amount of mass 
available for star formation towards the high density end 
is a key ingredient in the scenario of turbulence regulated 
star formation (e.g., Elmegreen \& Scalo 2004; Krumholz \& McKee 
2005).

To gauge the effect of the mass distribution on the star formation rate, 
we plot in Fig.~\ref{SFRff}(b) the SFR in the limit that all mass above 
a given density $\rho$ is converted into stars in one free fall time 
at the average density of that mass. The limiting rate is normalized 
by the global free-fall rate defined in equation~(1), which in our 
simulation has a value of 
\begin{equation}
{\dot M}_{\rm ff} = 4.32\times 10^2\ {c_s^3\over G},
\end{equation}
much larger than the classical value of $0.975 c_s^3/G$ for the 
inside-out collapse of a singular isothermal sphere (Shu 1977). The 
distribution of the free-fall rate has a broad peak around $0.4~\rho_0$. 
At lower densities, it increases with density, as a result of mass 
clumping into moderately overdense regions which, by itself, 
worsens the problem of rapid star 
formation. At densities above the peak, the mass fraction drops 
off with density faster than inverse square root of the density, 
leading to a decline of the free fall rate. For comparison, we 
note that the actual rate of star formation in the simulation 
is $\sim 0.03 {\dot M}_{\rm ff}$, or ${\dot M}_* \approx 13\ 
c_s^3/G$. 

\begin{figure}
\epsscale{0.8}
\plotone{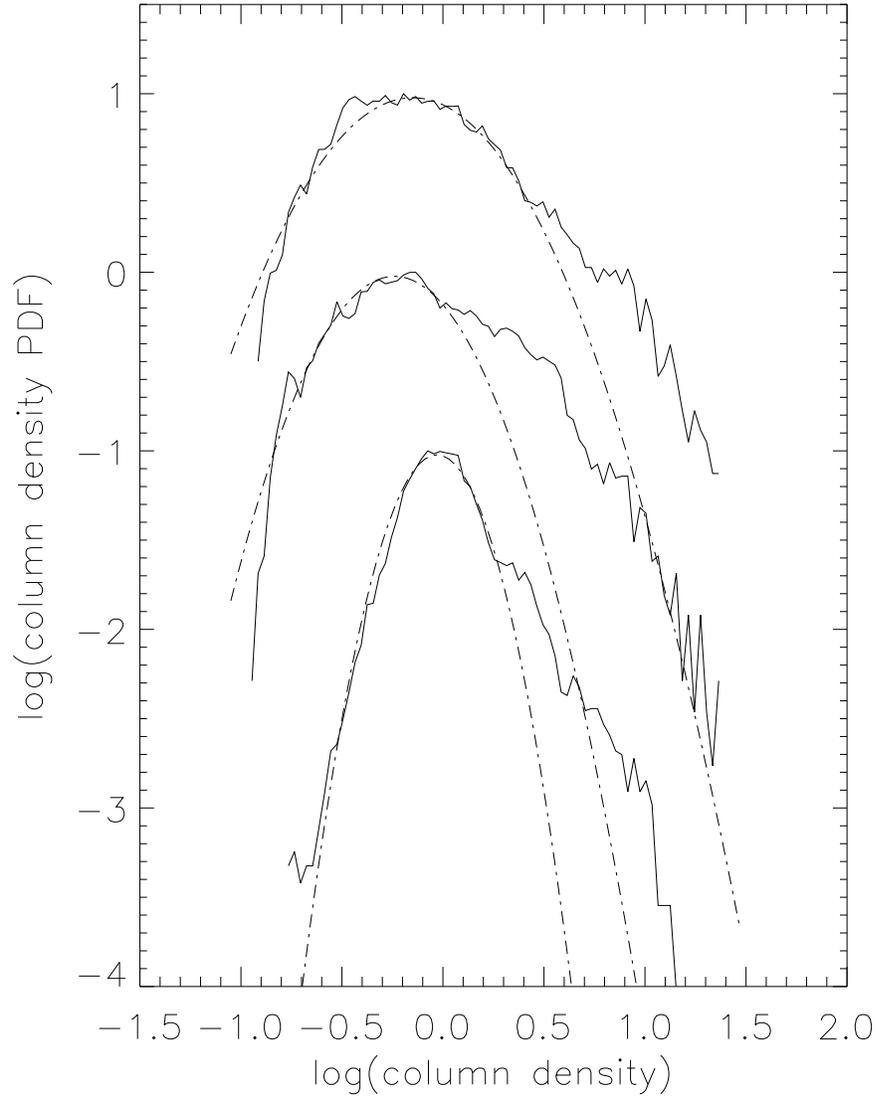}
\caption{Probability distribution functions (PDFs) of the column density 
(in units of $\rho_0 L_J$) along the $x$- (bottom), $y$- (middle) and 
$z$-axis (top). The curves are offset for clarity, with fitted lognormal 
distributions (dash-dotted lines) superposed for comparison.}
\label{coldenPDF}
\epsscale{1.0}  
\end{figure}

We next consider the distribution of the column density, which is 
more accessible to direct measurement than the volume density. 
In Fig.~\ref{coldenPDF}, we plot the column density PDFs at 
the representative time along the three axes. The 
PDFs along the $y$- and $z$-axes are broader than that along 
the $x$-axis. The broader distributions are caused by mass 
settling along the field lines; when viewed perpendicular to
the field lines, the column density is enhanced in the plane 
of mass concentration and decreased away from it. The PDFs 
deviate significantly 
from lognormal distributions, especially towards the high 
density end. The deviation is larger than those found by, for 
example, Ostriker et al. (2001) for decaying turbulence. 
One difference is that our turbulence is driven, indeed in 
a specific way---by collimated outflows. Perhaps more 
importantly, our prescription of star formation enables 
us to run the simulation longer, which allows more time for 
the cloud material to condense gravitationally into cores and 
filaments. In addition, the global gravitational potential well 
is deeper in our simulation. The cluster-wide gravity tends to 
concentrate dense cores and filaments towards the bottom of the 
potential well, increasing their chance for overlap, especially 
when viewed along the plane of mass concentration perpendicular 
to the large scale magnetic field. The overlap skews the PDF 
towards the high column density end.

\begin{figure}
\plotone{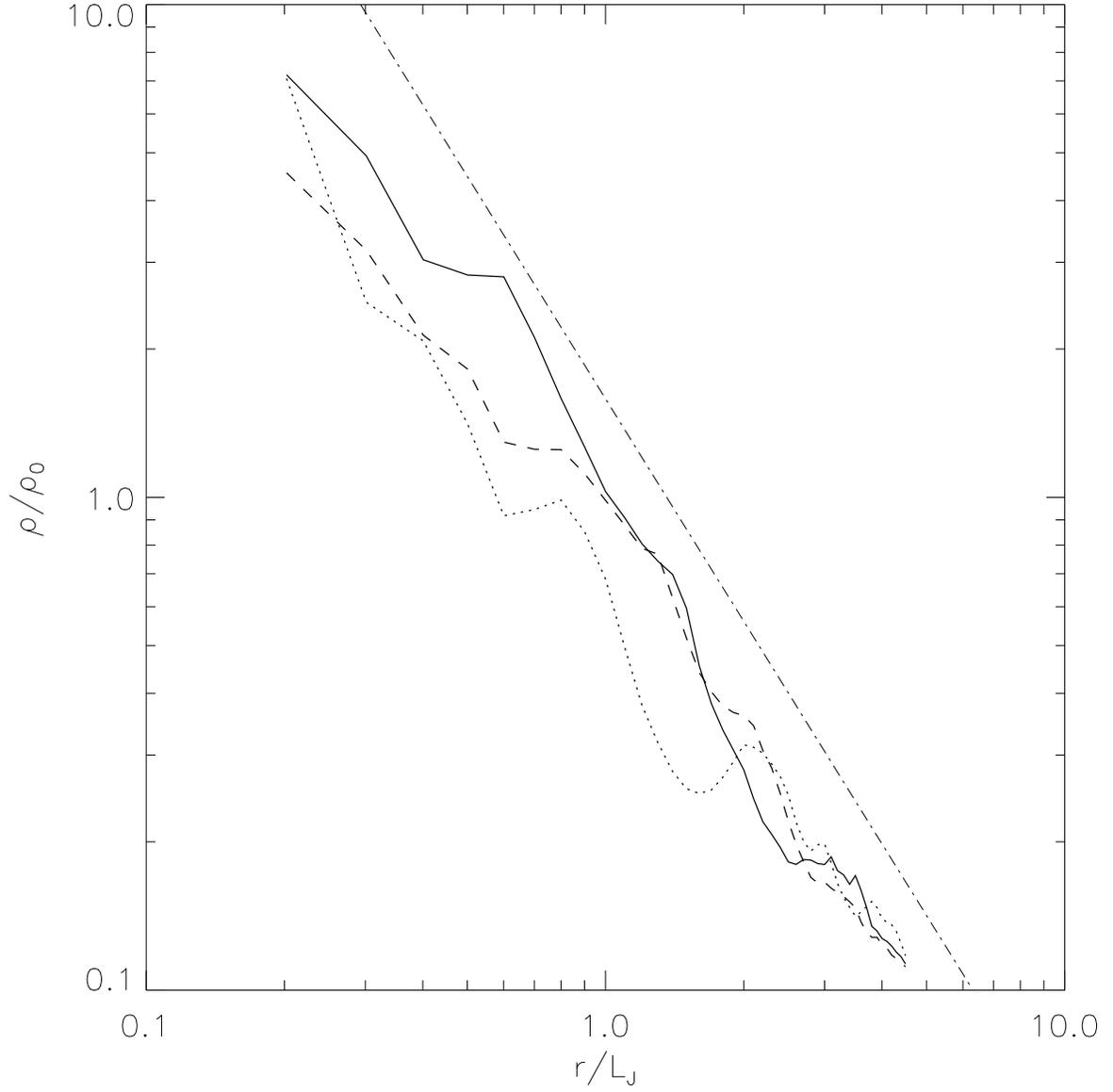}
\caption{Distributions of the spherically averaged density as a function 
of radius at the representative times t=1.0 (dashed), 1.5 (solid) and 
2.0~$t_g$ (dotted). A power-law $\rho \propto r^{-1.5}$ (dash-dotted 
line) is shown for comparison. }  
\label{Dprofile} 
\end{figure}

To examine the spatial distribution of the mass more quantitatively, 
we plot in Fig.~\ref{Dprofile} the average density as a function 
of radius at the representative time. The averaging is done in 
concentric shells centered on the location of minimum gravitational 
potential. We exclude in 
the log-log plot the central part of the cloud where the number 
of grid cells is small and the shell radius is ill-defined. 
There is a clear trend for the averaged density outside the
central (excluded) region to drop off with radius, in an 
approximately power-law fashion. 
A similar drop-off is found at the two other times (1.0 and 
2.0~$t_g$) shown in Fig.~\ref{Dprofile}. A power-law $\rho\propto 
r^{-1.5}$ is also plotted for comparison. Although the 
power-law provides a fair (although not unique) description of 
the averaged density distribution, it should be kept in mind 
that the medium is very clumpy. The clumpiness is illustrated 
vividly in Fig.~\ref{velvector}, where the volume density on a 
representative slice through the computational domain is 
displayed.

\subsection{Velocity Power Spectrum}

\begin{figure}
\epsscale{0.9}
\plotone{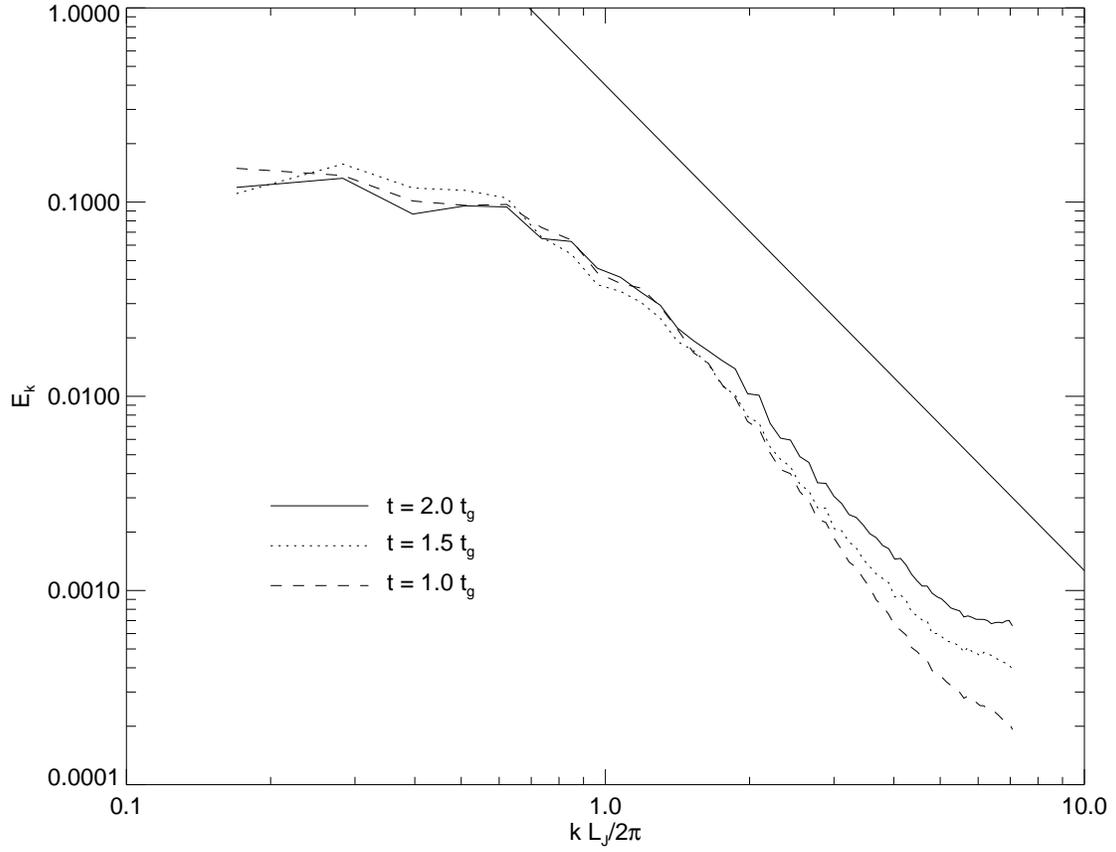}
\caption{Normalized velocity power spectra $E_k= k^2 v_k^2$ of the 
protostellar turbulence at three representative times 1.0 (dashed), 
1.5 (solid) and $2.0~t_g$ (dotted). Also shown for comparison is a 
power-law $E_k\propto k^{-2.5}$ (straight solid line).}
\label{spectra}  
\epsscale{1.0}
\end{figure}

The power spectrum of a turbulence in the inertial range (between the 
energy input and dissipation scales) can often be approximated by a 
power law, $E_k= k^2 
v_k^2 \propto k^{-n}$, where $k$ is the wavenumber. The 
power index $n$ holds clues to the nature of the turbulence. For 
example, the incompressible Kolmogorov turbulence has $n=5/3$. For 
the shock-dominated Burgers turbulence, the index is $n=2$. Some 
simulations of driven turbulence obtained a power index $n\approx 
1.74$ for isothermal gas (Boldyrev et al. 2002), which is closer 
to the Kolmogorov than Burgers value. The exact value of the power 
index depends, however, on model parameters, particularly the 
degree of cloud magnetization (Vestuto et al. 2003). In our standard 
simulation, the power spectra deviate strongly from a single 
power-law. The deviation is illustrated in Fig.~\ref{spectra}, where 
the spectra at three times (1.0, 1.5 and 2.0~$t_g$) 
are plotted. They all appear to have a break around $k_b \approx 
0.6\; (2 \pi/L_J)$. Below $k_b$, the spectra are rather flat. They 
steepen to approximately $k^{-2.5}$ above the break. The shape of 
the spectra indicates that the bulk of the power resides near the 
break rather than at the smallest wavenumber.

The break $k_b$ in the power spectrum corresponds to a characteristic 
length scale $L_b=2\pi k_b^{-1} \approx 1.7 L_J$, which is about 1/5 
of the size of the simulation box. The scale $L_b$ can plausibly be 
identified with the typical outflow length $L_f$. A crude estimate 
of $L_f$ comes from the distance that the (conical) jet component of 
the outflow travels before it is slowed down to the ambient turbulent 
speed ($v_{\rm turb} \approx 5 c_s$). Assuming a constant ambient 
density at the average value ($0.15 \rho_0$), we obtain $L_f \approx 
2.8 L_J$, which is about 65\% larger than $L_b$. However, the stars 
are formed preferentially in dense regions near the bottom of the 
potential well. The higher-than-average ambient density should 
lower the estimated outflow length somewhat, bringing it to a 
closer agreement with the characteristic break length. The outflow 
may be further shortened by magnetic tension if it propagates 
perpendicular to the large scale magnetic field. 
These considerations lead us to believe that the break in the 
power spectrum is produced by the momentum injection from 
protostellar outflows, although the strong inhomogeneity in the 
mass distribution makes it difficult to predict the length of 
any individual outflow accurately. 
Some outflows may be trapped close 
to where they are produced, while others may cross the entire 
cloud through largely empty regions. Matzner (2007) independently 
argued for the existence of a break in the power spectrum using 
a similar reasoning. 

Breaks are also seen in the power spectra of the turbulence driven 
in Fourier space. For example, Vestuto et al. (2003) drove their 
turbulence with a power peaking at a scale that is 1/8 of the 
box size. They found a break near the driving wavenumber. In the
inertia range above the break, they found that the power index 
$n$ decreases with increasing magnetic field strength. Our standard 
simulation has a field 
strength comparable to their model B, which has an estimated power 
index $n=2.0$. This index is smaller than that in our simulation. 
A potential cause for the difference is that their model B has a 
resolution higher than ours by a factor of 2, and Vestuto et al. 
have shown that a higher resolution tends to yield a smaller index. 
However, our preliminary analysis of a higher resolution ($256^3$) 
simulation indicates that in the inertial range the power index 
is essentially the same as that of our current simulation. The 
difference may instead result from the fact that self-gravity is 
included in our simulations but not in theirs and, perhaps more
importantly, that our turbulence is driven by discrete, highly
anisotropic outflows rather than isotropically in Fourier space. 
Anisotropy due to the collimation of individual outflows and 
the correlation of outflow directions is an important feature 
that distinguishes our turbulence simulations from the 
others and this feature is usually not taken into account in 
Kolmogorov-type dimensional analyses;
the effects of anisotropic driving remain to be fully explored.  
The difference in turbulence driving is also reflected in 
the flattening of our power spectra at the highest wavenumbers 
(see Fig.~\ref{spectra}). The flattening is most likely produced 
by the spherical component of the outflow, which supplies energy 
on small scales. 

The relatively flat power spectrum at small wavenumbers below the 
break $k_b$ is more interesting. Some of the power may come from 
inverse cascade. However, the inverse cascade in the simulations 
of Vestuto et al. (2003) produced an $E_k$ below the break 
wavenumber that decreases quickly towards small $k$ roughly as 
$E_k \propto k^2$. In contrast, the $E_k$ in our simulation 
generally remains flat or continues to increase slowly towards 
small $k$. We believe that 
most of the power below the break $k_b$ is supplied directly by 
collimated outflows rather than through inverse cascade. Movies 
of the standard simulation show clearly that many outflows can 
break out from the dense cores surrounding their driving sources 
and propagate to large distances, sometimes 
across the entire simulation box, injecting energy and momentum 
on the largest possible scale. Since the outflow propagation is 
well resolved in our simulations, we believe that the general 
conclusion that the power spectrum $E_k$ flattens at small 
wavenumbers is robust, although the details may depend on the 
outflow treatment and may be affected by the periodic boundary 
condition. 

The power spectrum of the protostellar turbulence can be decomposed 
into a solenoidal and compressible component. We find that the 
solenoidal component always dominates the compressible component. 
This is in qualitative agreement with previous simulations. 
Quantitatively, the solenoidal component is larger by a factor 
of $\sim 10$ at wavenumbers above the break $k_b$. Below the 
break, the factor is somewhat lower. Overall, the ratio of the 
two components is comparable to those found by Boldyrev et al. 
(2002) and Vestuto et al. (2003), who drove their turbulence by 
a purely solenoidal velocity field. The similarity indicates that 
the driving of protostellar turbulence by collimated outflows 
is probably mostly solenoidal, as one may expect from the large 
velocity shear between the fast outflow and the ambient medium. 

\subsection{Stellar Component}
\label{stars}

The simple prescription adopted in our model for star formation 
makes detailed comparison between the predicted stellar properties 
and observations premature at this stage of model development. 
Nevertheless, there are a few general features of the model stars 
that are worth noting. These include the spatial distribution 
of the stars relative to the gas, and their velocity dispersion. 

\begin{figure}
\plotone{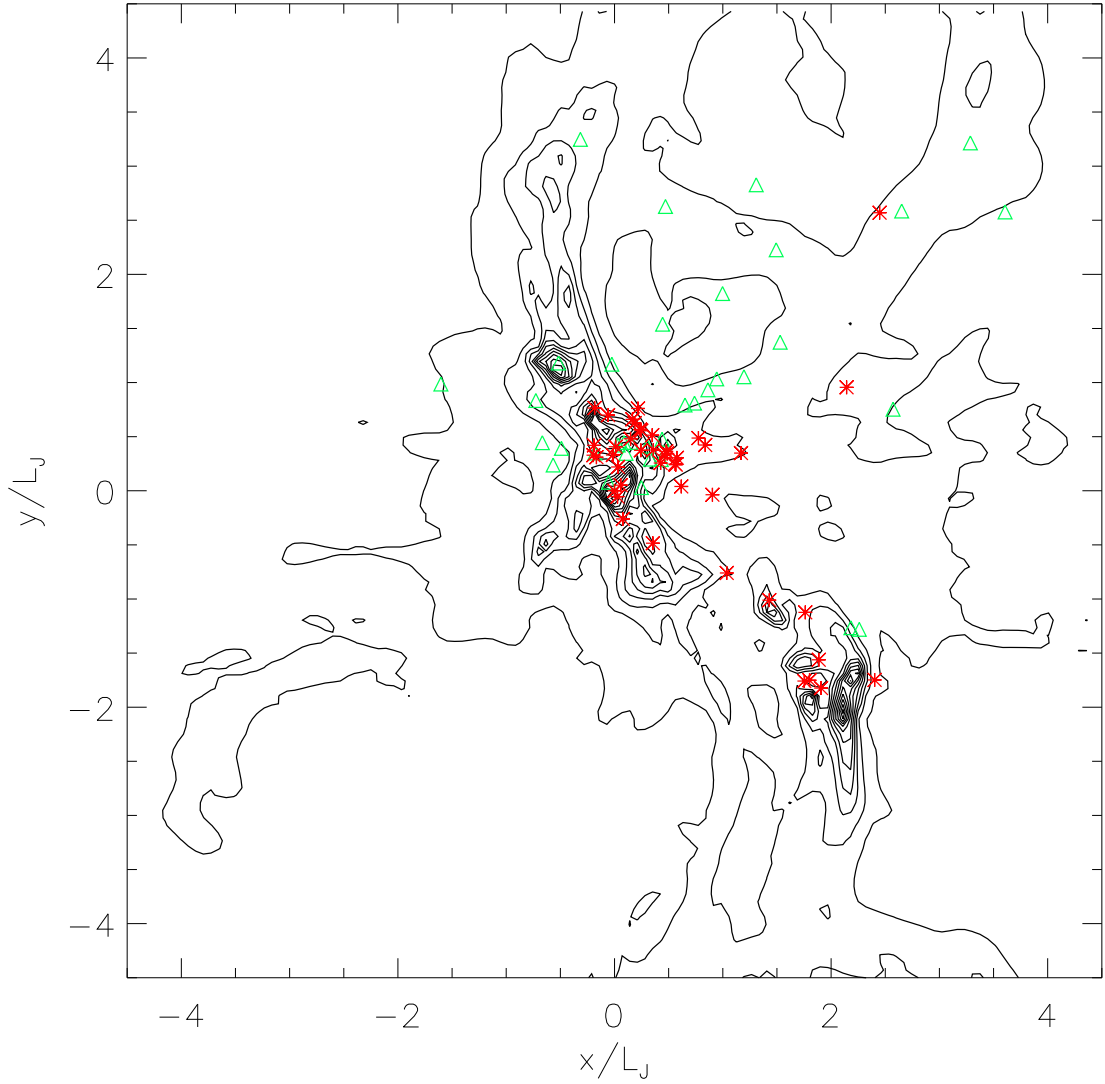}
\caption{Spatial distributions of the stars and gas at the 
representative time $1.5~t_g$. Stars formed before $1~t_g$ 
are denoted by green 
triangles, and those after $1~t_g$ by red asterisks. The contours 
are for the column density in units of $\rho_0 L_J$, starting 
from 1 and increasing in increments of 2.} 
\label{starmap}  
\end{figure}

We illustrate the stellar distribution in Fig.~\ref{starmap}, where 
the positions of the 80 stars at the representative time $1.5~t_g$ 
are projected onto the $x$-$y$ plane. Also plotted in the figure are 
contours of the column density along the $z$-axis, which delineate 
the gas distribution. 
The densest regions appear to form an ``S-shaped'' ridge. 
The elongation is caused mainly by mass 
settling along the large-scale magnetic field lines, which run 
more or less horizontally in the plot. There are two 
separate concentrations of dense gas, with the main one near 
the center of the column density map, and the secondary one 
to the lower right. The majority of stars are clustered around 
the main gas concentration. Specifically, slightly more than half 
of all stars (41 out of 80) are located within a radius $0.75~L_J$ 
(or 0.125~pc for the fiducial parameters) of the center. In this 
localized region, the SFE is $15.7\%$, much higher than that of 
the clump as a whole ($4.2\%$). The less massive concentration 
of dense gas is forming a smaller group of stars.

The degree of stellar clustering is expected to depend on age. To 
show the age dependence, we divide the stars into two groups: 
those born before and after $1~t_g$. They have 34 and 46 members 
respectively. The two groups of stars are represented by two 
different symbols and colors in Fig.~\ref{starmap}. It is clear 
that the younger stars are more closely associated with the dense 
gas at the current time and the older stars are more spread 
out, as expected. Part of the reason for the older stars to be 
more widely dispersed is that they have had more time to spread 
out. Another part is that the older stars are bound less tightly 
to the cluster-forming region as a whole. Indeed, 12 of the 34 
older stars (or 35\%) have positive total (kinetic minus 
gravitational) energies, and are thus formally unbound to 
the cluster at the time under consideration. The unbound fraction 
goes down to 8.7\% (or 4 out of 46) for the younger stars. The 
difference in the unbound fraction comes primarily from the 
tendency for the older stars to locate higher up in the potential 
well. As a result, they are able to move around more freely. If 
the gas in the cluster-forming region that gravitationally binds 
the majority of the stars together were to be removed suddenly, 
the stellar cluster would dissolve quickly. 

\begin{figure}
\plotone{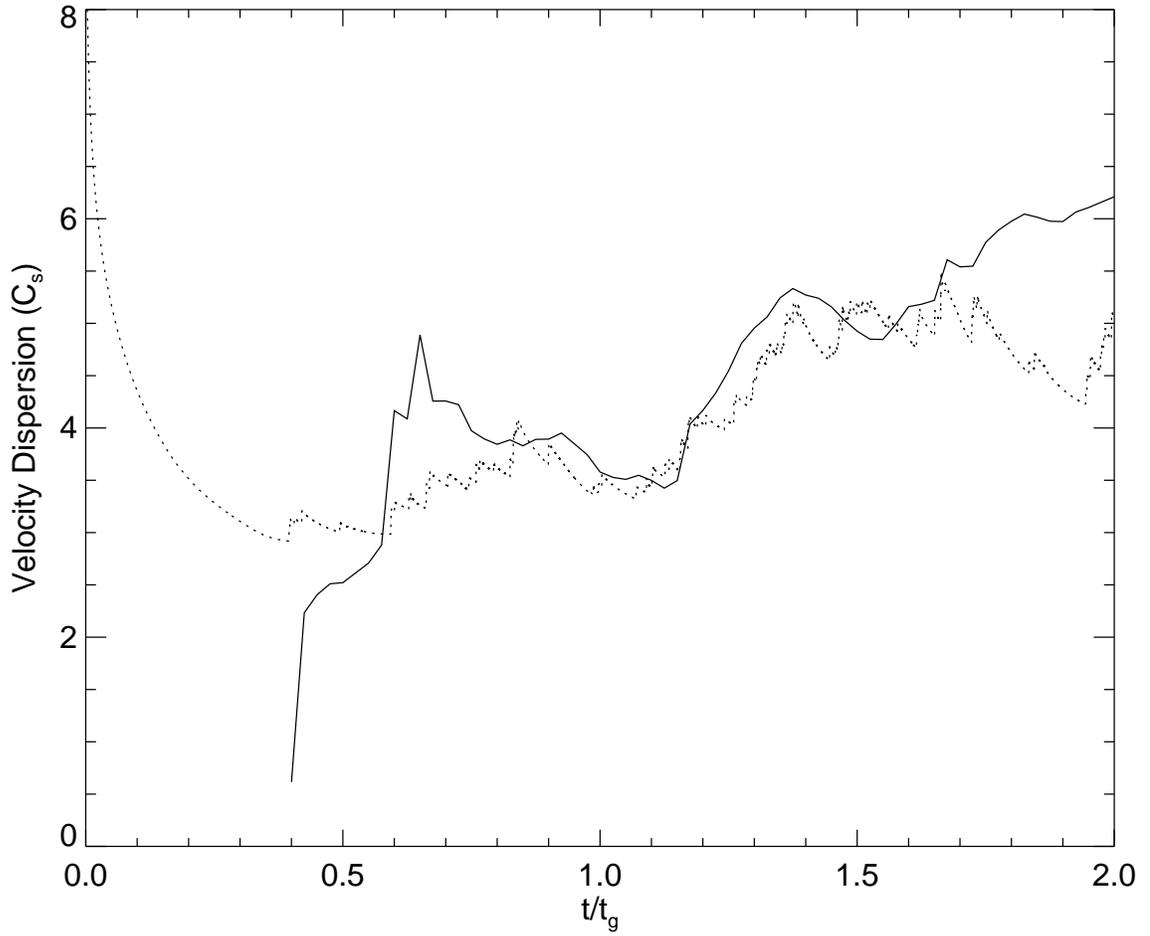}
\caption{Evolution of the stellar velocity dispersion $\sigma_*$ 
(solid line). Also plotted for comparison is the mass-weighted 
turbulent speed of the gas $v_{\rm turb}$ (dotted).}
\label{dispersion}  
\end{figure}

An important quantity that can in principle be directly measurable 
is the velocity dispersion of the stars. In Fig.~\ref{dispersion},
we plot the velocity dispersion $\sigma_*$, defined as the rms 
value of the stellar velocities relative to the mean\footnote{In 
computing $\sigma_*$, we exclude the two stars that move at 
speeds of 81.6 and 135~$c_s$, respectively, much faster than the 
other stars. Such fast moving stars are produced, on rare 
occasions, when the density in a fast moving outflow is pushed 
above the threshold for mass extraction. They may disappear with 
a refined outflow treatment.}, as a function of time for the 
standard model. Except for a brief initial period (when the number 
of stars is still small), the dispersion $\sigma_*$ ranges from 
$\sim 4$ to $\sim 6$~$c_s$. It is comparable to the mass-weighted 
average turbulent speed $v_{\rm turb}$ of the gas, which is also 
shown in the figure for comparison. 
Even though 
$\sigma_*$ and $v_{\rm turb}$ are defined somewhat differently, the 
fact that they are comparable is significant: both speeds reflect 
the depth of the global gravitational potential well. In particular, 
there is a significant increase in both 
$\sigma_*$ and $v_{\rm turb}$ starting around $1~t_g$. Their increase 
appears to track the increase in the absolute value of the 
gravitational energy around the same time (see Fig.~\ref{gravenergy}).

Before leaving the section on the standard model, we comment 
briefly on the dense cores, a number of which show up prominently 
in Fig.~\ref{starmap} (see also Fig.~\ref{velvector}). These 
cores are the ``crown jewels'' of the protostellar 
turbulence---the basic units of individual star formation. 
They tend to cluster near the bottom of the potential well.  
Whether they are created mostly in-situ near the bottom is 
unclear; some of them could be produced higher up in the potential 
well and later ``sink'' to the bottom. In any case, the spatial 
concentration of dense cores increases the chance for core-core 
interaction, particularly coalescence, which can build up the 
core mass. One the other hand, the mass of a core may be limited 
by gravitational collapse and disruption through outflows 
associated with star formation. The resulting core mass spectrum 
may hold the key to the determination of IMF (Motte et al. 1998). 
We will postpone a detailed investigation of these core-related
topics (including the role of outflow triggering) to future, higher 
resolution studies. 

\section{Variations on the Standard Model}

\subsection{Spherical versus Collimated Outflows}

Protostellar outflows are collimated, particularly during the earliest,
Class 0 phase of (low-mass) star formation. It is during this phase 
that the bulk of the mass of a star is assembled. If the 
outflows are driven by the gravitational energy release associated 
with mass accretion, then the bulk of their energy and momentum is 
expected to be injected into the ambient medium during this phase. 
This expectation is consistent with the observations of Bontemps 
et al. (1996), which showed that molecular outflows---the ambient 
material set into motion by the outflow-ambient interaction---are 
most powerful for the Class 0 sources; they tend to become weaker 
and broader with time (Bachiller \& Tafalla 1999). It is the momentum 
carried in the powerful molecular outflows in the early stages of 
star formation that we seek to capture with our outflow prescription
(see the Appendix); 
the optical jets observed at later times (Bally \& Reipurth 2001) 
and possibly the more tenuous wide-angle winds that are predicted 
to surround the jets in the magnetocentrifugal wind theory (Shu et 
al. 1995) are not included explicitly in our current model; their 
inclusion would only make the feedback from star formation stronger. 

\begin{figure}
\plotone{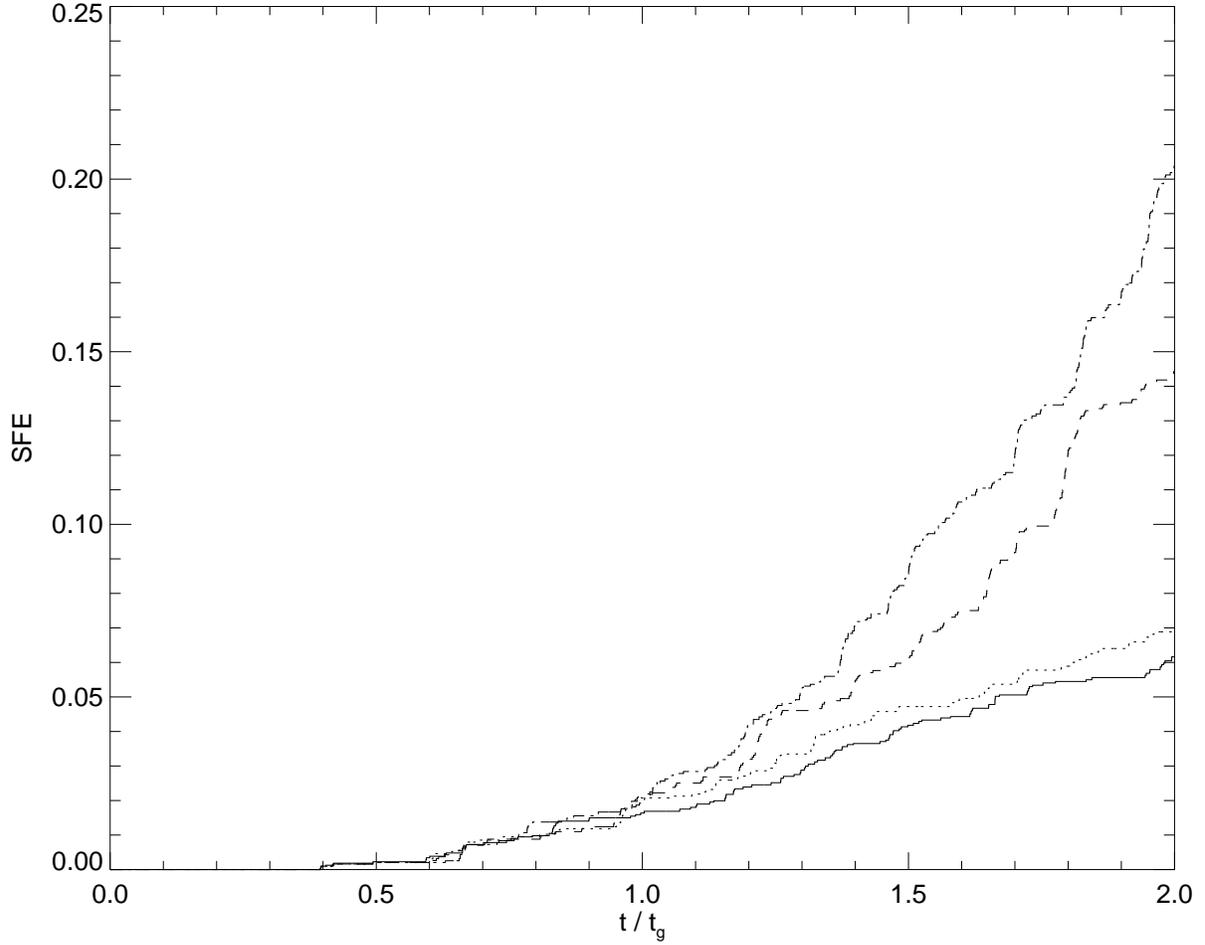}
\caption{SFEs for collimated and spherical outflow models. The 
upper most curve is for the spherical model (dash-dotted, Model 
E1), while the lower curves are for models of collimated outflows 
with a jet momentum fraction $\eta=0.25$ (dashed, Model E2), 
0.50 (dotted, Model E3) and 0.75 (solid, standard Model S0) 
respectively.}  
\label{SFEf05}
\end{figure}

We illustrate the effects of outflow collimation using three variants 
of the standard model. These variants include a 
spherical model that does not have a jet component at all (Model 
E1 in Table 1), and two models with weaker jet components (Model 
E2 with a jet momentum fraction $\eta=0.25$, and Model E3 with 
$\eta=0.50$). Other quantities, including the dimensionless
parameter $f$ for outflow strength, are kept the same as in the 
standard model. In Fig.~\ref{SFEf05}, we plot the time
evolution of the SFE for all four models. The SFEs remain 
relatively close together until about $1.0~t_g$. Between $1.0$ 
and $2.0~t_g$, there is a clear trend that the SFE decreases 
as the strength of the jet component increases relative to the 
spherical component. In particular, the SFE increases from 
$\sim 0.02$ to $\sim 0.20$ in the spherical model and to only 
$\sim 0.06$ in the standard jet model. The difference between 
the two in the total mass of the stars formed during this time 
interval is a factor of $\sim 4.5$. We conclude that collimated 
outflows are more efficient in suppressing star formation in 
the protostellar turbulence than the spherical outflow carrying 
the same amount of momentum. 

\begin{figure}
\epsscale{0.8}
\plotone{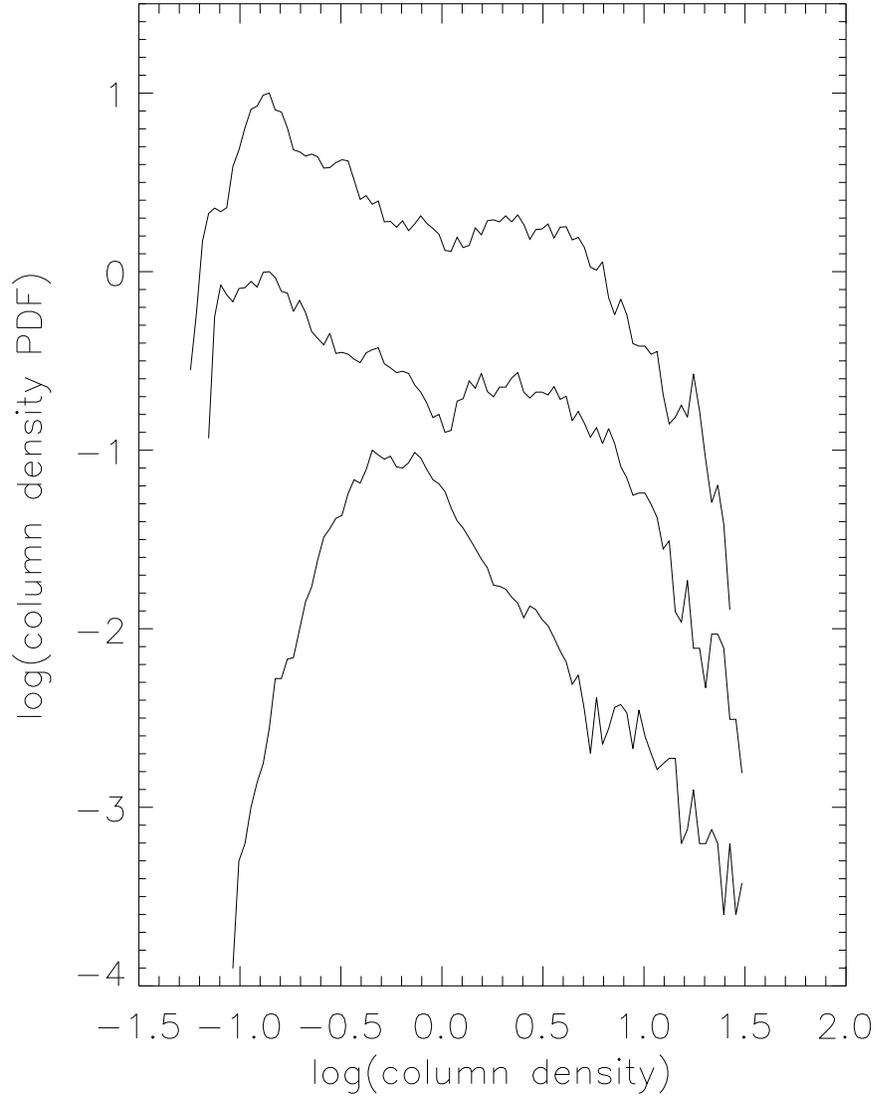}
\caption{The same as Fig.~\ref{coldenPDF} but for the spherical outflow 
model (Model E2) at the time $2 t_g$. }
\label{coldenPDF_sph}
\epsscale{1.0}  
\end{figure}

The above difference in SFE can be understood in terms of where the 
outflow momentum is deposited. Collimated outflows can propagate 
well outside the dense regions near the bottom of the potential 
well, where most stars form. They deposit a large fraction of their 
momenta in the outer envelope, where most mass resides. Spherical 
outflows, on the other hand, are more easily trapped. They tend 
to drive turbulent motions on a smaller scale, which are dissipated 
more quickly. The higher rate of turbulence dissipation in the 
spherical outflow model is compensated by a faster turbulence 
replenishment, through a higher rate of star formation. To fuel 
the higher star formation rate, more material must be concentrated 
into dense regions that can collapse to form stars. The mass 
concentration 
is reflected in the PDFs of the volume and column densities, both 
of which tend to be strongly skewed towards the high value end 
compared with lognormal distributions. The column density PDF is 
particularly noteworthy. In addition to a main peak similar to that 
shown in Fig.~\ref{coldenPDF} for the standard jet case, it often 
displays a broad shoulder or even a second peak to the right of 
the main peak, as illustrated in Fig.~\ref{coldenPDF_sph}. Such 
strong deviations from a lognormal distribution may be detectable 
in regions of active cluster formation when the outflows are 
trapped relatively close to their driving sources. 

\subsection{Effects of Outflow Strength}
\label{magstrength}

\begin{figure}
\plotone{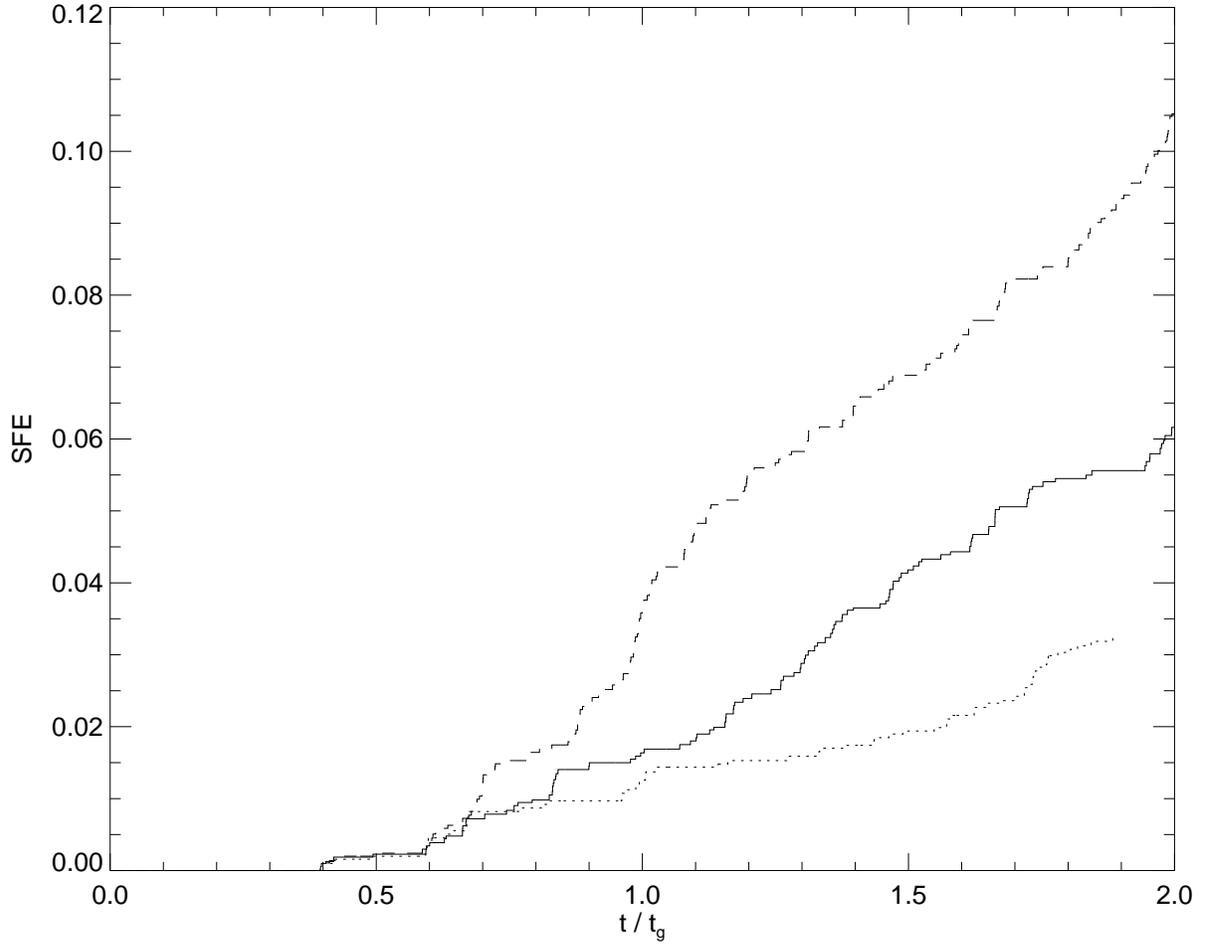}
\caption{Upper panel: SFEs for models of different outflow strengths. 
The curves are for the cases with outflow parameter $f=0.25$ (dashed, 
Model F1), $0.50$ (solid, standard Model S0) and $0.75$ (dotted, 
Model F2) respectively. Lower panel: Products of SFE and $f$, showing
that the total momentum injected into the cloud is insensitive to
the outflow strength.}  
\label{SFEf}
\end{figure}

The exact value of the outflow parameter $f$ is uncertain. In the 
Appendix, we have estimated a plausible range of $P_*=
10-100$~km/s for the outflow momentum per solar mass of stellar 
material, corresponding to $f=0.1-1$. To illustrate the effects 
of outflow strength, we have rerun the standard simulation but 
with a lower ($f=0.25$, Model F1) and higher ($f=0.75$, Model 
F2) outflow strength. The SFEs of these simulations are plotted in 
the upper panel of Fig.~\ref{SFEf}. There is a clear trend for 
the SFE to increase 
with decreasing outflow strength. Indeed, the SFE is nearly inversely
proportional to the outflow strength parameter $f$, so that their 
product is roughly the same for all three cases, as shown in 
the lower panel of the figure. 
Apparently, the total amount of momentum injected into the cloud 
is insensitive 
to the strength of individual outflow: the reduction in the momentum 
supplied per outflow is more or less compensated by the increase in 
the number of stars (and thus outflows). In contrast, the total 
amount of kinetic energy injected increases roughly linearly with 
the outflow strength. The difference is consistent with the 
expectation that the momentum of protostellar outflow is more 
directly relevant for turbulence replenishment than the energy; 
the latter dissipates more readily.

\begin{figure}
\plotone{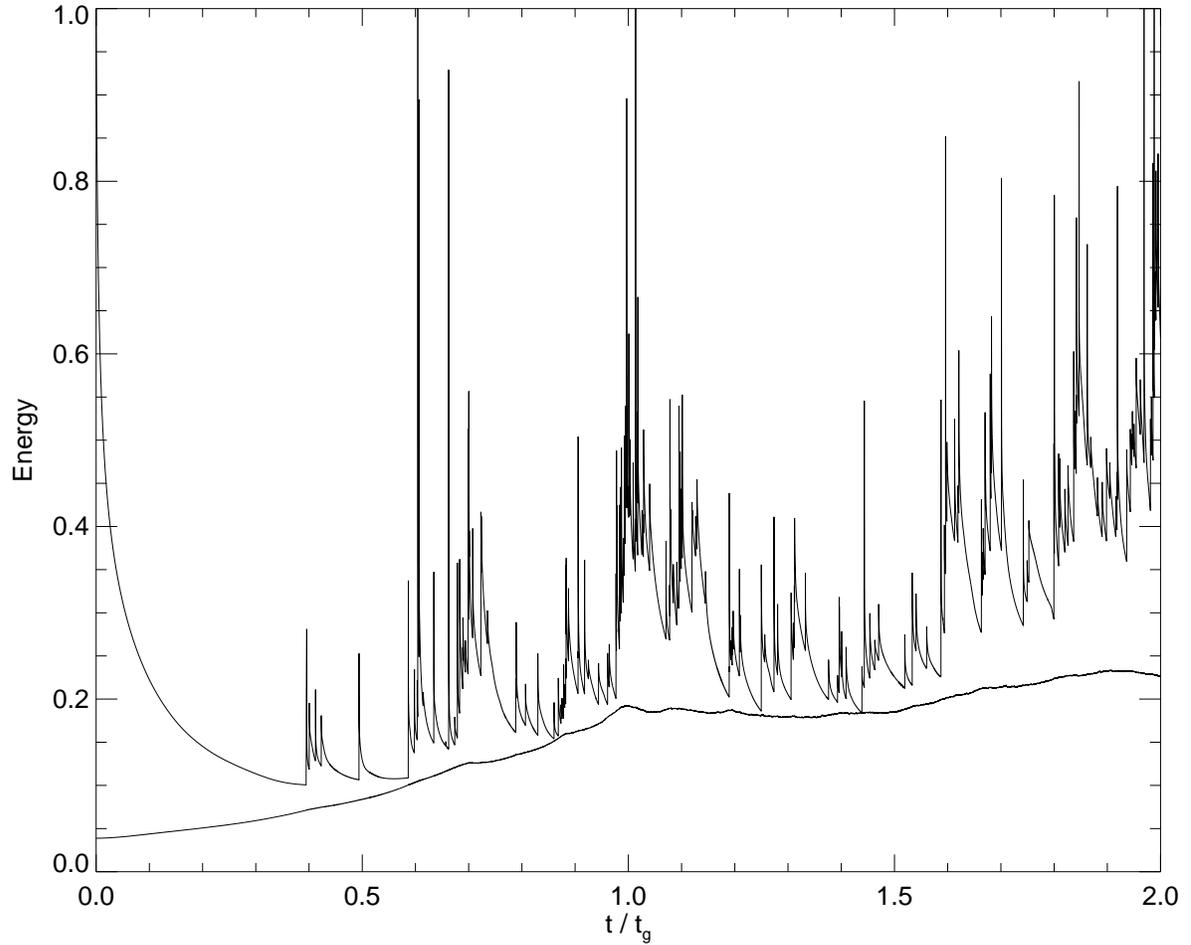}
\caption{Evolution of the kinetic and gravitational energies of 
the gas in Model F1 with relatively weak outflows ($f=0.25$). 
The energies are normalized by the initial kinetic energy of 
the turbulence, which is $50~c_s^2$ per unit mass, corresponding 
to a mass-weighted rms Mach number of 10.}  
\label{energyf025}
\end{figure}

Model F1 with $f=0.25$ is particularly interesting. It demonstrates 
that even relatively weak outflows are capable of driving a robust 
protostellar turbulence in which the SFR is reduced well below the 
limiting free-fall rate. In the reminder of the subsection, we will 
discuss this model in some detail, and contrast
it with the standard model, starting  with the energy evolution. As 
shown in Fig.~\ref{energyf025}, the total kinetic energy drops 
quickly, before being pumped up by the outflows associated with 
the (bursty) star formation. Compared with the standard model, the 
kinetic energy here
is more comparable with the gravitational energy, because it is less 
dominated by the (weaker) active outflows. The absolute value of the 
gravitational energy is somewhat higher than that in the standard
model, indicating that the bulk of the cloud material is more tightly 
bound. It increases initially, as a result of gravitational settling 
of the mass in the cluster-forming region towards the bottom of the 
potential well due to turbulence dissipation. The increase is stopped
around $1~t_g$ when enough motions are generated by outflows to arrest 
further contraction. There is some undulation at later times, which 
is also evident in the standard model (see Fig.~\ref{gravenergy}). The 
existence of such mild oscillations in the gravitational energy may 
not be too surprising, given the bursty nature of the star formation. 
The amplitude of the oscillation is small, again indicating that the 
cluster-forming system is hovering close to an equilibrium.

We have examined the quasi-equilibrium state of the weaker outflow 
model at the representative time $1.5~t_g$, and found it similar 
to that of the standard model discussed in \S~\ref{outflowsection}. 
In particular, the turbulent velocity field contains many distinct 
patches of more or less coherent motions. The average infall and 
outflow speeds are $-1.86$ and $2.57\ c_s$, respectively. The net 
average radial speed is only $1.09\times 10^{-2}\ c_s$, much smaller 
than the average infall and outflow speeds and the sound speed. The 
infall and outflow momenta nearly balance each other, with the 
slower infall speed compensated by a larger amount of infalling 
mass, as in the standard model. The mass weighted PDF of the 
volume density can again be fitted reasonably well with a lognormal 
distribution. The PDFs of the column density deviate more strongly 
from lognormal, especially along directions perpendicular to the 
initial magnetic field direction. A prominent break is present in 
the velocity power spectrum, as in the standard case. The 
similarities lead us to conclude that the gross properties of the 
protostellar turbulence are insensitive to the outflow strength. 

\subsection{Effects of Magnetic Field}
\label{magfield}

\begin{figure}
\plotone{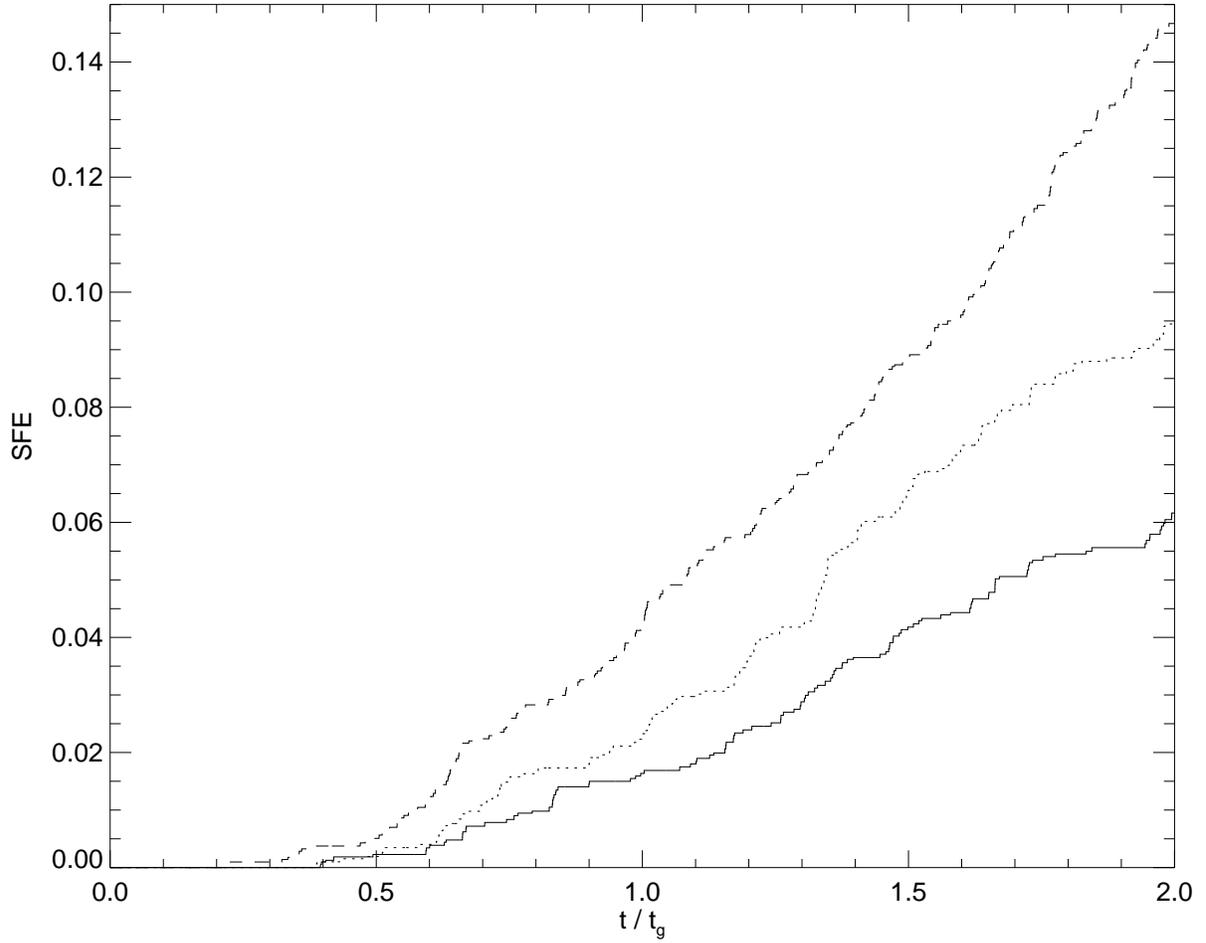}
\caption{SFEs for models of different magnetic field strengths. The 
curves are for the cases with $\alpha=10^{-6}$ (dashed, Model M1), 
0.5 (dotted, Model M2) and 2.5 (solid, standard Model S0) 
respectively.}   
\label{mag}
\end{figure}

We illustrate the effects of the magnetic field using three models 
of different field strengths. They are Models M1, M2, and S0 in 
Table~1, specified by $\alpha=10^{-6}$, $0.5$ and $2.5$, 
respectively. For the standard model ($\alpha=2.5$), the dimensionless 
mass-to-flux ratio ${\bar \Gamma}=0.52$ for the clump as a whole, as 
mentioned earlier. This ratio drops to $0.23$ and $3.3\times 10^{-4}$ 
for $\alpha=0.5$ and $10^{-6}$. In all three cases, the ratio is 
substantially less than unity, indicating that the global 
magnetic field is not strong enough to suppress star formation 
altogether. Stars are indeed formed in all three cases, as shown 
in Fig.~\ref{mag}, where the SFEs of the models are plotted. There 
is a clear trend for the SFE to decrease with increasing field 
strength, as one might expect. Specifically, the magnetic field in 
the standard model ($\alpha=2.5$) has reduced the rate of star 
formation by a factor of $\sim 2.2$ compared to the negligible 
field case ($\alpha=10^{-6}$). Even the relatively weak field in 
the $\alpha=0.5$ model appears to have a significant effect on 
the rate of star formation, reducing it by a factor of $\sim 1.6$.

A simple, albeit crude, way to gauge the dynamical importance of 
a magnetic field is to compare its energy to the kinetic energy 
of the gas. In our case, the 
comparison is complicated by the fact that the kinetic energy is 
often dominated by active outflows. For example, in the standard 
simulation with $\alpha=2.5$, $81\%$ of the energy is carried 
by $8.6\%$ of the mass that moves faster than $10\ c_s$ at the 
representative time $1.5~t_g$. The remaining energy is carried 
by the bulk, more slowly moving material. It has a value of $8.80
\times 10^2$ in units of $\rho_0 c_s^2 L_J^3$, corresponding to 
a mass-weighted rms speed of $4.55\ c_s$, comparable to the 
specific (scalar) momentum (see Fig.~\ref{momentum}). This portion 
of kinetic energy is smaller than the total magnetic energy at the 
same time ($2.63\times 10^3\ 
\rho_0 c_s^2 L_J^3$). However, the magnetic energy is dominated by 
the background uniform field, which accounts for $1.82\times 10^3\ 
\rho_0 c_s^2 L_J^3$. The remaining $8.06\times 10^2\ \rho_0 c_s^2 
L_J^3$, carried by the distorted magnetic field, is remarkably close 
to the kinetic energy carried by the bulk of the turbulent material 
away from the active outflows ($8.80 \times 10^2\ \rho_0 c_s^2 
L_J^3$). The similarity indicates that an energy equipartition 
is reached between the distorted magnetic field and the turbulent 
motions of the bulk material for the standard model, which has a 
moderately strong magnetic field to begin with. 

\begin{figure}
\plotone{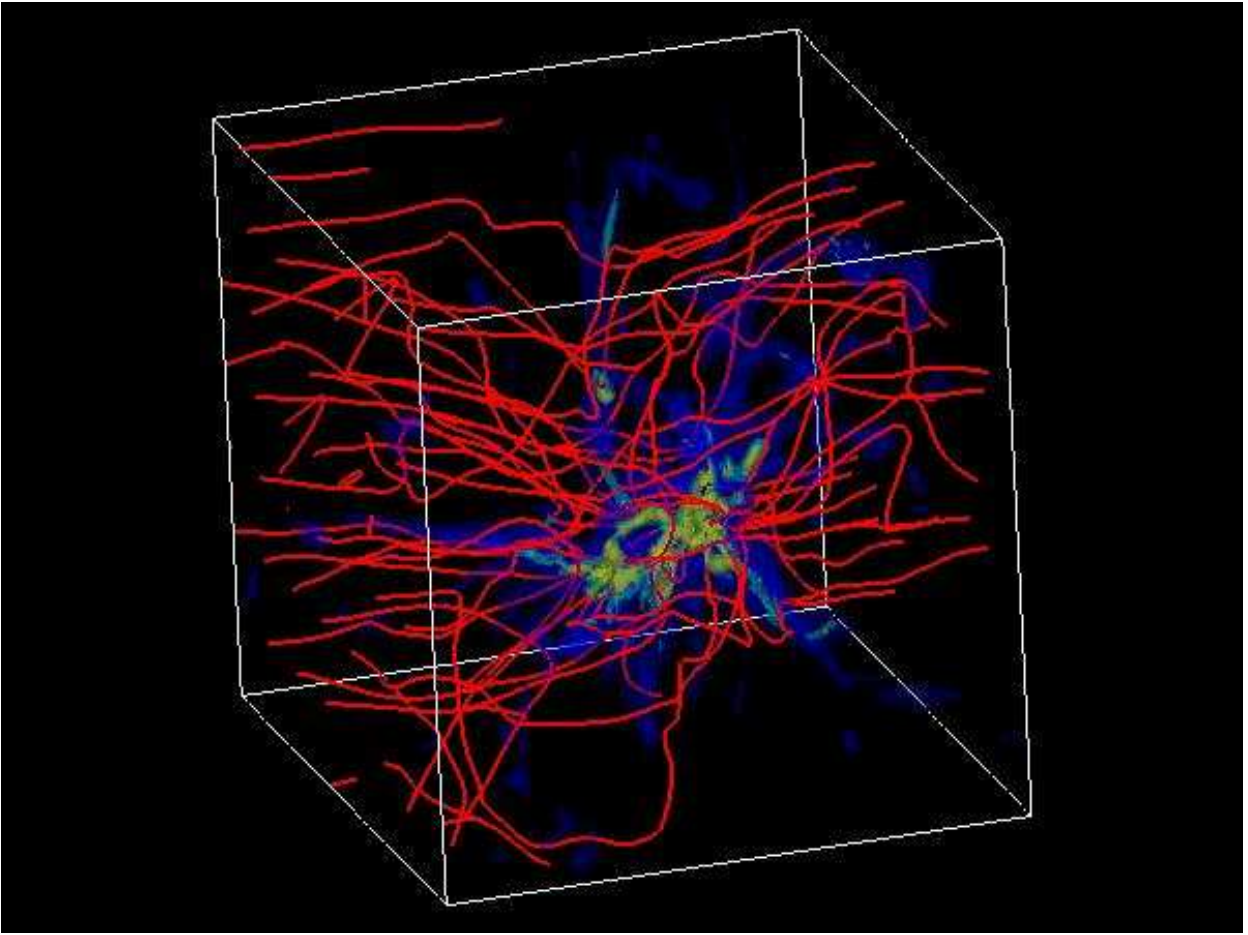}
\caption{Magnetic field lines and density distribution in 3D for 
Model M2 ($\alpha=0.5$). The field lines are drawn in red. The 
blue and yellow isodensity surfaces are for 1 and $30\ \rho_0$ 
respectively.}  
\label{3Dfield}
\end{figure}

The weaker field case of $\alpha=0.5$ (Model M2) is more intriguing. 
At the beginning of the simulation, the magnetic energy is only 
$3.65\times 10^2\ \rho_0 c_s^2 L_J^3$, well below the kinetic energy. 
By the representative time $1.5~t_g$, it has nearly quadrupled to 
$1.38\times 10^3\ \rho_0 c_s^2 L_J^3$. Most of this energy is stored
in the distorted magnetic field, which accounts for $1.02\times 10^3 
\rho_0 c_s^2 L_J^3$. The distorted field energy is higher than its 
counterpart in the standard model by $27\%$, despite the fact that 
the standard model is more strongly magnetized to begin with. This
energy is 
again close to the kinetic energy carried by the bulk of the cloud 
material 
that moves at a speed below $10\ c_s$ ($1.11\times 10^3\ \rho_0 c_s^2 
L_J^3$). Apparently, the magnetic energy has been amplified to an 
equipartition level in this initially weaker field case. Part of 
the amplification comes from the concentration of mass towards the 
bottom of the potential well, which drags the field lines into 
pinched configurations. The pinched field lines are 
evident in Fig.~\ref{3Dfield}, which shows the field structure and 
isodensity surfaces in 3D. More dramatic amplification comes from 
the stretching of field lines by fast moving outflows, which creates 
large magnetic distortions that are relatively long lived. Further 
amplification comes from the turbulent motions of the (slower) 
bulk material. These processes remain to be quantified.

The amplification factor is even larger for the weakest field model 
of $\alpha=10^{-6}$. The initial magnetic energy is $7.29\times 
10^{-4}\ \rho_0 c_s^2 L_J^3$. It increases to $0.46\ \rho_0 c_s^2 
L_J^3$ at $1.5~t_g$, by a factor of $630$. Despite the large 
enhancement factor, the magnetic energy is still orders of 
magnitude below the kinetic energy, indicating that the magnetic 
field is dynamically unimportant in this extreme case. We conclude 
that magnetic fields are dynamically important in protostellar 
turbulence as long as their strengths are not much below the 
critical value to begin with. The role of magnetic fields is
discussed further towards the end of \S~\ref{nature} below. 

\section{Discussion}

\subsection{The Nature of Protostellar Turbulence}
\label{nature}

A salient feature of protostellar turbulence is the simultaneous 
existence of infall and outflow motions. Fluid parcels are pushed 
up the gravitational potential well by outflows. Once slowed 
down, they are pulled back towards the bottom of the potential 
well by gravity, setting up a vigorous circulation of material 
between the dense central region where most stars form and 
the outer envelope, where most of the mass resides. The 
circulation is in a way reminiscent of convection, although 
the fluid motions are highly supersonic, unlike the conventional 
buoyancy-driven convection. In this picture, the gravity plays 
a role as important as the outflows. It drives infall motions 
that close the global circulation, enabling the system to reach 
a quasi-equilibrium state, in which the infall and outflow 
motions are globally balanced. Simultaneous infall and outflow
are observed in a number of embedded clusters, including NGC 
1333 (Walsh et al. 2006; Knee \& Sandell 2000), the Serpens 
cloud core (Williams \& Myers 2000; Olmi \& Testi 2002; Davis 
et al. 1999), and NGC 2264 (Peretto et al. 2006; Williams \& 
Garland 2002; Wolf-Chase et al. 2003). Such observations 
underpin our notion of cluster formation in outflow-driven 
protostellar turbulence. 

There are several lines of evidence for a quasi-equilibrium state in 
all of our simulations. First, the global infall and outflow momenta 
nearly cancel, yielding a net mass-weighted velocity in the radial 
direction much smaller than the average infall or outflow velocity 
overall. There are, however, regions where one dominates the 
other and the equilibrium is locally upset.
Second, both the specific momentum and gravitational energy 
approach, and oscillate with a small amplitude around, a 
constant value at late times, indicating that an equilibrium 
has been reached. The total kinetic energy is not as good an
indicator of equilibrium, since it is typically dominated by 
fast moving transients. The kinetic energy for the regions 
away from active outflows is, however, comparable to the 
gravitational energy, suggesting a rough virial equilibrium 
for the bulk material. In addition, the spherically averaged 
density distribution as a function of radius appears to settle 
into a power-law of roughly constant power index in the 
envelope. The density PDFs and velocity power spectra also 
maintain similar shapes at late times. All these lines of 
evidence point to the existence of a quasi-equilibrium state 
with fully developed protostellar turbulence; it is in such 
an environment that the majority of the cluster members form. 

Stars form at a relatively slow rate. In our standard model, 
only $\sim 3\%$ of the gas is converted into stars per free fall 
time at the average density. This corresponds to a remarkably long
gas depletion time of $\sim 33$ free fall times. The exact value 
of the star formation rate depends on several factors, including 
the strength and degree of collimation of the outflow, as well as 
the initial magnetic field strength (and perhaps topology, which
is not explored here). Although there is considerable 
uncertainty in estimating each of these factors, we believe that 
the general conclusion that outflows can maintain the turbulence in 
a cluster forming clump is robust. At a fundamental level, there is 
enough momentum in the outflows to replenish the momentum dissipated 
in the turbulence for a reasonable star formation efficiency of 
order $10\%$, as stressed by McKee (1989) and Shu et al. (1999), 
among others, as long as the momentum dissipation time is not much 
shorter than the free fall time. Our detailed simulations allowed 
us to estimate the dissipation time self-consistently. For the 
set of models with relatively strongly magnetic fields ($\alpha
=2.5$) and collimated outflows (Models S0, M1 and M2), the 
average dissipation time, as measured by the ratio of the 
equilibrium (scalar) momentum and the rate of momentum input, 
is close to the free-fall time ${\bar t}_{\rm ff}$ at the 
average density, more or less independent of the outflow strength. 
The dissipation time increases with the strength of the magnetic 
field and the degree of outflow collimation.

The velocity field of protostellar turbulence is dominated by the 
solenoidal (or shear) component. 
Such a velocity field is perhaps to be expected, given that there 
is a large shear between the collimated outflows and the ambient 
medium, and that any fast compressive motions that may have also 
been generated by the interaction are readily dissipated in shocks. 
Indeed, the 
large ratios of the kinetic energies in the solenoidal and 
compressive components obtained in our simulations are similar 
to those obtained in Boldyrev et al. (2002) and Vestuto et al. 
(2003), where the turbulence is driven isotropically
in Fourier space using 
a purely solenoidal velocity field. Our protostellar turbulence 
is a special case of driven turbulence: it is driven by discrete, 
fast-moving, collimated outflows (see also Mac Low 2000), that 
come from the protostars formed preferentially near the bottom 
of the potential well. For turbulence driven in Fourier space, 
the dissipation rate depends on the scale of driving (e.g., 
Mac Low 1999), with those driven on small scales decaying faster 
than those driven on large scales. 
A well recognized problem with driving in Fourier space is that 
it affects the gas in the entire computation box simultaneously 
(e.g., Elmegreen \& Scalo 2004). 
The more physical outflow-driving acts sequentially on a range 
of scales, from near the protostars to large distances, and 
anisotropically, both on the local scale of individual star
formation and on the global scale of cluster-forming clump;
the latter comes about because the outflow orientations are 
not completely random in the presence of a strong magnetic 
field. The amounts of energy and momentum deposited on 
different scales 
depend on both the intrinsic properties of the outflows (such 
as the flow speed, degree of collimation, duration, etc), and 
the (generally anisotropic) distributions of mass and magnetic 
field in both the dense cores that surround the outflow-driving 
protostars and the general turbulent background. Despite these 
differences, our finding that the turbulence driven by collimated 
outflows decays more slowly than that driven by spherical 
outflows is qualitatively consistent with the previous result, 
since collimation effectively increases the scale of driving.

Magnetic fields can influence the protostellar turbulence in
several ways. If strong enough, they can flatten the mass
distribution and affect, perhaps even control, the directions 
of outflow ejection and propogation. In our simulations, we
find that the energy stored in the distorted magnetic field is 
comparable to the kinetic energy of the bulk cloud material 
away from active outflows, as long as the field strength is 
not far below the critical value to begin with; relatively weak 
fields are amplified by gas motions to an equipartition value. 
An implication is that the turbulent motions for the bulk of 
the gas is roughly Alfv{\'e}nic, and the magnetic field is 
dynamically important. Indeed, the field may be crucial in  
transmitting the outflow energy and momentum to regions not 
directly impacted by outflows, through large amplitude Alfv{\'e}n
waves. It may also prevent dense fragments from moving freely in 
the clump potential, since they need to drag along the ambient 
material linked to them through magnetic field lines (Elmegreen 
2006). These effects remain to be fully quantified.

Approximate equipartition between the magnetic and turbulent kinetic 
energies is inferred in regions with Zeeman 
measurements of the magnetic field strength (e.g., Myers \& Goodman 
1988; Crutcher 1999). For the energy in the measured (ordered) 
magnetic field to be comparable to the turbulent kinetic energy, 
the flux-to-mass ratio probably needs to lie within a factor of a 
few of the critical value (as in the standard simulation). Such a 
ratio may be obtained 
naturally if the cluster-forming clumps are created quickly out of 
a more diffuse, magnetically subcritical medium 
through strong 
shocks, driven perhaps by HII regions or supernova explosions. The
strong C-shock may reduce the flux-to-mass ratio in the shocked
layer below the critical value, inducing a relatively rapid 
subsequent evolution (collapse and star formation) that freezes 
the flux-to-mass ratio at a value somewhat below the critical 
value. In this picture of externally triggered clump formation 
out of a (perhaps moderately) magnetically subcritical cloud (to 
be quantified elsewhere), the magnetic energy would be automatically 
comparable to the turbulent kinetic energy, and be a significant 
factor in regulating star formation in clusters. 

To summarize, the protostellar turbulence in active regions of cluster 
formation is outflow-driven, gravity-assisted, and magnetically 
mediated. 

\subsection{Connection to Previous Work and Observations}

The present investigation is a first step towards a quantitative 
theory of cluster formation in outflow-driven protostellar 
turbulence. Our emphasis has been on the global properties of 
the turbulence. Supersonic turbulence in molecular clouds has 
been subject to intensive numerical studies in recent years 
(as reviewed, e.g., by Elmegreen \& Scalo 2004). A longstanding 
problem is that it decays away quickly, which prompted many 
workers to drive the turbulence in Fourier space. The most important 
distinction of our simulations is that we drive the turbulence 
in physical space, using outflows that are ubiquitously observed 
in star formation. 

There have been a few previous studies of turbulent motions driven
by outflows. Allen \& Shu (2000) examined the effects of outflows 
on the dynamics of critically magnetized, sheet-like GMCs. The
same 2D geometry is adopted in our previous investigation of star 
formation in turbulent, magnetically subcritical clouds that 
includes both ambipolar diffusion and outflows (Nakamura \& Li 
2005). These 2D calculations are extended in this paper to three 
dimensions (but without ambipolar diffusion). The only other 3D 
study is that of Mac Low (2000). He conducted a preliminary 
investigation of continuous driving of turbulence in a 
non-self-gravitating cloud by outflows ejected from a number of 
randomly distributed locations that are fixed in space and time. 
We went beyond this study by including self-gravity and connecting 
the outflows to the discrete events of star formation, which occur 
preferentially near the bottom of the potential well. 

%

The protostellar outflow has a unique feature that can be potentially 
observable. It has a velocity power spectrum characterized by a 
prominent break. At small wavenumbers, the spectrum remains 
relatively flat. It steepens to approximately $E_k\propto 
k^{-2.5}$ above the break. The broken power-law spectrum implies 
that the bulk of the energy resides near the break, rather than
at the smallest wavenumber. Given that the turbulence is driven 
by a collection of outflows of varying lengths, there is no
prior reason why the power should be dominated by the largest
scale of the system. In particular, the Larson's (1981) 
linewidth-size relation is unlikely to be strictly applicable 
{\it inside} parsec-scale clumps of active star formation, as 
recognized independently by Matzner (2007). High resolution 
observations of nearby embedded clusters that resolve the gas 
kinematics on different scales can be used to test this proposition
(e.g., Caselli \& Myers 1995; Saito et al. 2006). 

Another way to probe the nature of the turbulence is through 
probability distribution functions (PDFs). Although the PDFs of 
the volume density in protostellar turbulence can often, but not 
always, be fitted reasonably well by lognormal distributions, 
the PDFs of the column density show considerable deviations, 
especially towards the high value end and in regions of rapid
star formation. The strong deviation is 
not necessarily a unique signature of the protostellar turbulence 
per se. More likely, it is a reflection of the crucial role 
that the gravity plays in our simulations. The gravity 
allows dense cores to form and collapse individually, and to 
cluster near the bottom of the potential well collectively. 
The ability to follow the cloud evolution to a later stage of 
gravitational evolution distinguishes our simulations from other 
grid-based simulations of cluster formation in turbulent clouds 
(e.g., Heitsch et al. 2001; Ostriker et al. 2001; Li et al. 
2004; Tilley \& Pudritz 2004; Vazquez-Semadeni et al. 2005; for 
examples of SPH simulation of cluster formation, see Klessen et 
al. 1998 and Bate et al. 2003). 
Such simulations are typically terminated before the runaway 
collapse of the densest core, before the primordial turbulence 
is transformed into a protostellar turbulence through the outflows
associated with star formation. We are able to go well beyond 
the collapse of the first core and follow the formation of 
multiple generations of stars and their feedback into the 
cluster-forming environment, albeit using simple prescriptions 
for the subgrid physics. We hope to refine these prescriptions 
in the future. 

\subsection{Limitations and Future Refinements}
\label{improvement}

The most restrictive prescription is perhaps the mass of individual 
stars. We have on purpose kept the stellar mass close to 
a single value, about $0.5~M_\odot$ for the fiducial choices of
parameters, which is comparable to the typical mass of low mass
stars. The prescription, however, decoupled the properties of
the dense cores and the stars that they produce. This is 
obviously an oversimplification. A better prescription is 
perhaps to assume a constant accretion time for all stars. 
Some motivation for this prescription comes from the analysis
of Myers \& Fuller (1993), who inferred a spread in mass accretion 
time for stars between 0.3 and 30 M$_\odot$  much smaller than 
the spread in stellar mass. Ultimately, high resolution calculations, 
using perhaps adaptive mesh refinement (AMR), are required to 
follow the process of mass accretion in detail to determine 
the stellar mass self-consistently. Even with AMR, there are 
uncertainties of physical origin that cannot be eliminated with
higher resolution: the effects of jets and winds from young 
stellar objects (as well as radiation for massive stars), which 
can interfere with the process of mass accretion and limit the 
stellar mass (Nakano et al. 1995; Matzner \& McKee 2000; Shu 
et al. 2004).  

In our simulation, the momentum of the primary outflow from close to 
a star is assumed to be imparted instantaneously to a small region 
immediately next to the star. While we believe that this simple 
prescription captures the essence of the outflow feedback, a more 
realistic treatment would be to include the primary protostellar wind 
in the simulation for a finite duration. A potential difficulty with 
such a treatment is that the high wind speed may reduce the time step 
to the point of making the global simulation prohibitive. Another 
possibility is to carry out separate calculations of core collapse 
and outflow propagation (e.g., Shang et al. 2006), and find general
rules that can be used in the global calculations. In any case, 
a detailed treatment of outflow-ambient interaction is needed 
for not only the determination of the mass of an individual star
in a given core, but also the outflow-shaped environment for 
the cluster formation. The theoretical treatment can benefit
greatly from systematic observations that quantify the 
dependence of outflow properties on the stellar mass and 
star-forming environments, if any. There is a pressing need for 
such observations. 

Another area of future improvement will be in spatial resolution.
Higher resolution is desirable for at least two reasons. First, 
the protostellar turbulence appears to be dominated by shearing 
rather than compressible motions. The shear may excite 
small scale turbulence that may be washed out by numerical
diffusion on a relatively coarse grid, although the magnetic 
field may suppress the development of such a turbulence. 
More importantly, higher resolution will allow us to study 
the properties of dense cores with greater confidence. Dense 
cores in some cluster forming regions appear to have mass 
spectra similar to the IMF. Understanding this important 
observation will be a focus of our future higher resolution 
simulations. 

Lastly, we reiterate that the standard periodic boundary condition 
is adopted in our simulations. It has the undesirable effect of 
preventing energy and momentum, carried by either outflowing 
material or MHD waves, from leaving the simulated region. This 
effect may be compensated to some extent by the energy and 
momentum fed into the region from the ambient environment. A 
consequence of this restriction is that we are unable to 
follow cloud disruption by outflows, which may terminate the 
formation of relatively poor clusters that do not contain 
massive stars. 

\section{Summary}

We have carried out numerical simulations of cluster formation 
in protostellar outflow-driven turbulence. Our emphasis was on 
the global properties of the gas and the star formation rate. 
The main results are summarized below.

1. Protostellar outflows of strength in the observationally inferred 
range can readily replenish the supersonic turbulence in clumps of
active cluster formation against dissipation, and keep the clumps 
close to a dynamical equilibrium. The protostellar turbulence is 
characterized by the coexistence of infall driven by gravity and 
outflow motions. These motions are roughly balanced so as to yield a 
net mass flux towards the bottom of the gravitational potential well 
much smaller than that expected in a global free-fall collapse. 

2. The protostellar turbulence is maintained by outflows associated 
with star formation at rates as low as a few percent per free fall 
time. Collimated outflows are more efficient in driving the 
turbulence than spherical outflows carrying the same amount of
momentum. Collimation enables an outflow to propagate farther 
away from its source, effectively increasing the scale for energy 
and momentum injection. This result is in agreement with the 
previous finding that turbulence driven on a larger scale decays 
more slowly. Although protostellar outflows tend to retard 
global star formation, they can directly trigger star formation 
in localized regions through shock compression. 

3. There is a prominent break in the velocity power spectrum of 
the protostellar turbulence. Below the break wavenumber, the 
spectrum is relatively flat. It steepens at higher wavenumbers. 
The break may provide a handle to distinguish the outflow-driven
protostellar turbulence from other types of turbulence. In 
particular, it is unlikely that the Larson's linewidth-size holds 
on the scales of relatively flat power spectrum. Another general 
feature is that the turbulent velocity field is dominated 
by the solenoidal component, produced perhaps by the strong shear
between the outflow and the ambient medium. The high degree of 
anisotropy in the turbulence driving is an important feature of 
the protostellar turbulence that warrants further investigation. 

4. We find that the PDFs of the volume density in the cluster-forming
clump can be approximated reasonably well by lognormal distributions 
in general. The PDFs of the column density, on the other hand, often 
show large deviations from lognormal, especially when the ordered 
magnetic field is strong enough to flatten the mass distribution
along the field lines, and in regions of rapid star formation. Dense 
cores and filaments, which cluster near 
the bottom of the potential well, tend to skew the column density 
PDFs towards the high value end. The mass distribution in the cluster
forming region is very clumpy. Nevertheless, the spherically averaged 
density typically increases with decreasing radius in an approximately 
power-law fashion in the envelope, with $\rho\propto r^{-1.5}$ 
providing a reasonable fit to the profile in many cases.   

5. Magnetic fields are dynamically important for cluster formation even 
in magnetically supercritical clumps, as long as the initial field 
strength is not much below the critical value to begin with. Moderately 
strong fields can significantly reduce the rate of star formation. Even 
a relatively weak magnetic field can be amplified to an equipartition 
level by the outflow-driven turbulent motions. The magnetic field is 
expected to influence the directions of outflow ejection and propogation
and the transmission of outflow energy and momentum to the ambient 
medium, although these effects remain to be fully quantified. 

6. We find that the stellar velocity dispersion is comparable to the 
turbulent speed of the gas. More detailed predictions on the stellar 
properties await a refined treatment of the subgrid physics of 
individual star formation. 

\acknowledgments 
This work is supported in part by a Grant-in-Aid for Scientific Research 
of Japan (1540117), a Grant for Promotion of Niigata University Research 
Projects, and NSF and NASA grants (AST-0307368 and NAG5-12102). We thank 
Chris Matzner for sharing some results prior to publication and the
referee for comments that improved the presentation of the paper. The 
numerical calculations were carried out mainly on SX6 at Niigata University, 
on SX8 at YITP in Kyoto University, and on SR8000 at Center for Computational 
Science in University of Tsukuba.

\section{APPENDIX: Protostellar Outflow Momentum}
\label{outflow}

In this Appendix, we estimate a plausible range of 
the outflow momentum from several observations.
Most, perhaps all, low-mass stars go through a phase of strong outflow 
during formation. 
The origin of the outflow is unclear. The leading scenario is that they 
are driven by rotating magnetic fields from close to the central object 
(e.g., K\"onigl \& Pudritz 2000; Shu et al. 2000). These winds interact 
with the environments that surround the forming stars, feeding energy 
and momentum back into the star-forming cloud. The best evidence for 
the wind-ambient interaction comes from bipolar molecular outflows 
(Lada 1985), which are thought to be the cloud material that has been 
set into motion by the interaction. 
Since the interaction is likely momentum (rather than energy) 
conserving, a useful quantity for characterizing the strength of 
the feedback is the total (time-integrated) momentum of the wind 
divided by the stellar mass, denoted by $P_*$ (Matzner \& McKee 2000). 
Its value can be constrained from both observations 
and theoretical considerations.

The best observational constraint on the wind momentum per unit
stellar mass $P_*$ comes from the famous source of bipolar
molecular outflow, L1551 IRS5. The CO outflow is the first to 
be discovered (Snell et al. 1980), and remains arguably the 
best studied source (Stojimirovic et al. 2006). Just as 
importantly, the mass of its central object has been determined
dynamically, based on radio observations of the orbital motions
of the central binary system (Rodriguez et al. 2003); this mass 
is typically not available for other strong outflow sources, which 
are deeply embedded in general. 

Stojimirovic et al. (2006) carried out a thorough study of the 
structure and kinematics of the outflow, and obtained a flow 
momentum between $20.5 - 26.5\ M_\odot$ km~s$^{-1}$ (depending 
on the excitation temperature adopted), corrected for the
effects of CO optical depth but not inclination. The inclination 
angle $i$ of the outflow to the plane of the sky is somewhat 
uncertain. Fridlund \& Liseau (1994) found values  
ranging from $\sim 24^{\circ}$ to $\sim 43^{\circ}$ from 
the radial velocities and proper motions of HH knots that 
bisect the blue outflow lobe. Consistent 
with this range is the inclination angle ($i\sim 30^\circ$) for
the symmetry axis of the system inferred from the modeling of 
IR images in scattered light (Lucas 
\& Roche 1996) and continuum images of circumstellar disks at 
7 mm (Rodriguez et al. 1998). If we adopt the same range for 
the molecular outflow as well, the inclination-corrected momentum 
would increase to $\sim 30.1 - 65.2 M_\odot$ km~s$^{-1}$. 
This is likely a lower limit to the total momentum carried by 
the primary wind, because it does not include the contribution 
from the slowest part of the outflow (within 1.5 km/s of the
systematic velocity) that is difficult to disentangle from 
the ambient cloud and there is evidence that the wind may have 
punched through part of the visible boundary of the L1551 
cloud. Using the total stellar mass of $\sim 1.2$ M$_\odot$ 
for the binary estimated by Rodriguez et al. (2003), we find 
$P_* \gsim 25 - 54$ km s$^{-1}$ for this particular source. 
%
%

L1551 IRS5 is considered prototypical of moderately collimated, 
``classical'' molecular outflows (Bachiller \& Tafalla 1999).
There is another class of highly collimated outflows, driven 
mostly by Class 0 objects. One of the best studied outflows in this 
class is L1157, driven by a low luminosity protostar of 
$\sim 11 L_\odot$ (Umemoto et al. 1992). Bachiller et al. (2001) 
obtained a 
total flow momentum of $4.71 M_\odot$ km/s, after correcting for 
the effects of CO optical depth but not inclination. A large  
inclination correction factor is needed, since the outflow appears 
to lie close to the plane of the sky. If one adopts the value $i\sim 
9^\circ$ deduced by Gueth et al. (1996), the momentum would 
increase by a factor of $6.4$, to $\sim 30 M_\odot$ km/s. The
mass of the central object is uncertain. For a 
rough estimate, we assume that most of the bolometric luminosity 
$L_{\rm bol}$ comes from the accretion luminosity
\begin{equation}
L_{\rm acc} = {G {\dot M_*} M_* \over R_*}
\label{accretionL}
\end{equation}
and that the mass accretion rate ${\dot M_*}$ is given by $M_*/
t_{\rm dyn}$, where $M_*$ is the stellar mass and $t_{\rm dyn}$ 
is the dynamical time for the outflow, which is measured to be 
$\sim 3\times 10^4$ years for L1157 (Bachiller et al. 2001). 
Under these assumptions, we find 
\begin{equation}
M_* \approx \left({L_{\rm bol} R_* t_{\rm dyn} \over G}\right)^{1/2}
=0.17 \left({L_{\rm bol}\over 10 L_\odot}\right)^{1/2} \left({ 
R_*\over 3 R_\odot}\right)^{1/2} \left({t_{\rm dyn}\over 3\times
10^4 {\rm yrs} }\right)^{1/2} M_\odot,
\label{stellarM}
\end{equation} 
where the radius $R_*$ of the (low-mass) protostar should be within 
a factor of 2 of the scaling $3 R_\odot$ (Stahler 1988). Froebrich 
et al. (2003) estimated a somewhat lower stellar mass ($0.1 M_
\odot$) for L1157, adopting a more elaborate, time dependent model 
for mass accretion and a lower $L_{\rm bol}=7.6\pm 0.8 L_\odot$. These 
crude estimates, if not too far off the true value, would point to 
a large value for the wind momentum per unit stellar mass $P_*$ of 
more than $10^2$ km/s for this particular case.

The above values of $P_*$ estimated for the best studied cases may 
not be representative, however. 
They are likely biased toward high values because of 
observational selection effects; weaker outflows are more 
difficult to study in detail. To account for the possibility of a 
range of values for $P_*$, we introduce a dimensionless parameter 
\begin{equation}
f= \left({P_* \over 100\ {\rm km/s} }\right) = \left({V_{\rm w}
\over 10^2\ {\rm km/s}}\right) \left({M_{\rm w}\over M_*}\right).
\label{windpara}
\end{equation}
It is the product of the speed of the primary wind $V_{\rm w}$ in units 
of $10^2$~km/s and the fraction of the stellar mass that is ejected in 
the wind. For revealed T Tauri stars where the wind speed can be 
measured directly, 
$V_{\rm w}$ is typically a few hundred km/s. In the X-wind theory 
(Shu et al. 2000), the most natural value for the ratio $M_{\rm w}/M_*$
is $\sim 1/3$. It is therefore conceivable that $f$ may be as high as 
unity. Some support for a relatively high value of $f$ comes from the 
unpublished work of Cabrit \& Shepherd 
(quoted in Richer et al. 2000), which indicates that $(M_{\rm w} 
V_{\rm w}) / (M_* V_{\rm K}) \sim 0.3$ (where $V_{\rm K}=[GM_*/R_*]
^{1/2}$ is the Keplerian speed at the stellar surface) over a wide 
range of stellar luminosity. 
For a typical young star of mass $M_* \approx 0.5 M_\odot$ and radius $R_*
\approx 3 R_\odot$, we have $P_* \sim 0.3\ V_{\rm K}\approx 53$~km/s (or
$f\approx 0.5$. 
On the other hand, the wind during the deeply embedded phase (when most 
of the driving of molecular outflows occurs) may be somewhat slower,  
both because of a smaller stellar mass (and thus a shallower gravitational
potential well) and because of a likely higher wind mass loading which, 
in the magnetocentrifugal wind theory, would lead to a slower outflow 
(e.g., Anderson et al. 2005). Also, there is evidence 
that the ratio of mass loss rate in the wind and accretion rate onto 
the star ${\dot M}_{\rm w}/{\dot M}_* \approx 0.1$ during both the revealed 
(Calvet 1997) and the embedded phase (Bontemps et al. 1996). These 
estimates, although fairly uncertain, point to values for $f$ as low 
as $\sim 0.1$.

\begin{deluxetable}{llllll}
\tablecolumns{6}
\tablecaption{Model Parameters \label{tab:model}}
\tablewidth{\columnwidth}
\tablehead{
  \colhead{Model}     & \colhead{$\alpha$} & \colhead{$f$}
 &\colhead{$\eta$}   &\colhead{Note} 
}
\startdata
S0   & 2.5 & 0.5  & 0.75  & standard model, with jet component\\
E1   & 2.5 & 0.5  & n/a   & spherical outflow, no jet component \\
E2   & 2.5 & 0.5  & 0.25   & significantly weaker jet component\\
E3   & 2.5 & 0.5  & 0.50   & somewhat weaker jet component\\
F1   & 2.5 & 0.25  & 0.75 & weaker outflow\\
F2   & 2.5 & 0.75  & 0.75 & stronger outflow\\
M1   & $10^{-6}$ & 0.50  & 0.75 & extremely weak magnetic field\\
M2  & 0.5 & 0.5  & 0.75 & weaker magnetic field \\

\enddata
\tablecomments{The quantity $\alpha$ is the ratio of magnetic 
to thermal pressure at the cloud center, $f$ the dimensionless 
parameter for outflow strength, and $\eta$ the fraction of 
outflow momentum carried in the jet component. All models are
run to $2~t_g$, except for Model F2, which is terminated around 
$1.8~t_g$ because the time step becomes prohibitively small.}
\end{deluxetable}

\end{document}